\documentclass[12pt]{article}

\usepackage{amsmath,amsfonts,amssymb}
\usepackage{mathptmx}
\usepackage{txfonts}
\usepackage{lineno}

\usepackage{graphicx}

\usepackage[letterpaper,margin=1in]{geometry}

\usepackage[nolist,nohyperlinks]{acronym} %
\usepackage[colorlinks=true,unicode,linkcolor=blue,citecolor=cyan,]{hyperref}

\renewenvironment{abstract}
	{\quotation}
	{\endquotation}

\date{}

\makeatletter
\renewcommand{\fnum@figure}{\textbf{Figure \thefigure}}
\renewcommand{\fnum@table}{\textbf{Table \thetable}}
\makeatother

\usepackage{scicite}

\usepackage{url}

\def\scititle{
The role of the solid–melt interface in accelerating the self-catalyzed growth kinetics of III–V semiconductors
}
\title{\bfseries \boldmath \scititle}

\author{
        Zhucong~Xi$^{1,\dagger}$,
        Abby~Liu$^{1,\dagger}$,
        Xiaobo~Chen$^{2}$,
        Meng~Li$^{2}$,
        Dmitri~N.~Zakharov$^{2}$,\and
        Judith~C.~Yang$^{2,3,4}$,
        Rachel~S.~Goldman$^{1,5,6}$,
        Liang~Qi$^{1,\ast}$\and
	\small$^{1}$Department of Materials Science and Engineering, University of Michigan, Ann Arbor, MI 48109, USA.\and
        \small$^{2}$Center for Functional Nanomaterials, Brookhaven National Laboratory, Upton, NY 11973, USA.\and
        \small$^{3}$Department of Chemical and Petroleum Engineering, University of Pittsburgh, Pittsburgh, PA 15261, USA.\and
        \small$^{4}$Department of Physics and Astronomy, University of Pittsburgh, Pittsburgh, PA 15261, USA.\and
        \small$^{5}$Department of Physics, University of Michigan, Ann Arbor, MI 48109, USA.\and
        \small$^{6}$Applied Physics Program, University of Michigan, Ann Arbor, MI 48109, USA.\and
	\small$^\ast$Corresponding author. Email: qiliang@umich.edu\and
	\small$^\dagger$These authors contributed equally to this work.
    }

\begin{document} 
\begin{acronym}
    \acro{BEP}{Brønsted-Evans-Polanyi}
    \acro{CI-NEB}{climbing image nudged elastic band}
    \acro{CTR}{crystal truncation rod}
    \acro{DFT}{density functional theory}
    \acro{KMC}{kinetic Monte Carlo}
    \acro{KRA}{kinetically resolved activation}
    \acro{VLS}{vapor-liquid-solid}
    \acro{VS}{vapor-solid}
    \acro{MC}{Monte Carlo}
    \acro{MD}{molecular dynamics}
    \acro{MLIP}{machine learning interatomic potentials}
    \acro{MBE}{molecular-beam epitaxy}
    \acro{MEP}{minimum energy path}
    \acro{NEB}{nudged elastic band}
    \acro{STM}{scanning tunneling microscopy}
    \acro{STEM}{scanning transmission electron microscopy}
    \acro{TEM}{transmission electron microscopy}
    \acro{VASP}{Vienna \textit{Ab Initio} Simulation Package}
    \acro{NW}{nanowire}
    \acro{WZ}{wurtzite}
    \acro{ZB}{zinc blende}
    \acro{MLIP}{machine learning interatomic potential}
    \acro{AIMD}{\textit{ab initio} molecular dynamics}
    \acro{LPE}{liquid phase epitaxy}
    \acro{ES}{Ehrlich-Schwoebel}
    \acro{CV}{collective variable}
    \acro{PAW}{projector-augmented wave potentials}
    \acro{PBE}{Perdew-Burke-Ernzerhof}
    \acro{VTST}{Transition States Tools}
    \acro{LAMMPS}{Large-scale Atomic/Molecular Massively Parallel Simulator}
    \acro{PFP}{PreFerred Potential}
    \acro{MSD}{mean-squared displacement}
    \acro{HAADF-STEM}{high-angle annular dark-field scanning transmission electron microscopy}
    \acro{HRTEM}{high-resolution transmission electron microscopy}
    \acro{STEM}{scanning transmission electron microscopy}
    \acro{FEL}{free energy landscape}
\end{acronym}

\maketitle



\begin{abstract}  
\bfseries \boldmath
Solid-melt interfaces play a pivotal role in governing crystal growth and metal-mediated epitaxy of gallium nitride (GaN) and other semiconductor materials. Using atomistic simulations based on machine-learning interatomic potentials (MLIPs), we uncover that multiple layers of Ga atoms at the GaN-Ga melt interface form structurally ordered and electronically charged configurations that are critical for the growth kinetics of GaN. These ordered layers modulate the free energy landscape (FEL) for N adsorption and substantially reduce the migration barriers for N at the interface compared to a clean GaN surface. Leveraging these interfacial energetics, kinetic Monte Carlo (KMC) simulations reveal that GaN growth follows a diffusion-controlled, layer-by-layer mechanism, with the FEL for N adsorption emerging as the rate-limiting factor. By incorporating facet-specific FELs and the diffusivity/solubility of N in Ga melt, we develop a predictive, fitting-free transport model that estimates facet-dependent growth rates in the range of $\sim$0.01 to 0.04~nm/s, in agreement with experimental growth rates observed in GaN nanoparticles synthesized by Ga-mediated molecular beam epitaxy (MBE). This multiscale framework offers a generalizable and quantitative approach to link atomic-scale ordering and interfacial energetics to macroscopic phenomena, providing actionable insights for the rational design of metal-mediated epitaxial processes.
\end{abstract}

\section*{Teaser}
A unified multiscale model predicts GaN and III-nitride epitaxy growth kinetics from solid-melt interface energetics.

\section*{Introduction}
Understanding solid-melt interfaces is crucial to numerous technologically important processes, ranging from bulk crystal growth~\cite{burton1951growth,magnussen2019atomicscale} to liquid metal-mediated epitaxy growth~\cite{chiritescu2007ultralow}. Prime examples of the critical role of the solid-melt interface are the \ac{LPE}~\cite{moustakas2011invited,moustakas2012role,jaramillo-cabanzo2019liquid} and \ac{VLS} mechanism~\cite{colombo2008gaassisted,guniat2019vapor}, both enabling controlled morphology and high crystalline quality when fabricating semiconductors. During \ac{LPE} or self-catalyzed \ac{VLS} growth of III-V semiconductors~\cite{mandl2010growth}, a liquid metal droplet, composed of the group III element (e.g., liquid Ga or In), serves both as the reservoir and as a catalyst. This liquid captures group V atoms (e.g., N or As) from the vapor phase and facilitates their delivery and incorporation into the growing solid crystal~\cite{wang2007realtime}. The successful incorporation of these atoms into the crystal lattice is therefore greatly mediated by the solid-melt interface. 

Although some macroscopic parameters, such as growth temperatures, vapor partial pressures~\cite{reyes2013unified,fernandez-garrido2009growth,lu2021influence}, and droplet contact angle~\cite{jacobsson2016interface} have been used to interpret and control epitaxy, they generally reflect the underlying atomistic processes at the solid-melt interface that govern growth kinetics. The atomistic-level interactions and structural arrangements at interfaces dictate the thermodynamics and kinetics, as well as the ultimate macroscopic properties of the resulting materials. For instance, temperatures and pressures determine the chemical potentials of the species involved~\cite{glas2013predictive} and surface diffusion barriers~\cite{kaufmann2016critical}. In addition, the contact angles of the droplets in \ac{VLS} growth are indicators of the interfacial energies, which are closely related to the composition of the droplet~\cite{oliveira2021role,glas2007why}. These atomistic factors critically influence the atomic incorporation processes at the interface, ultimately affecting morphology, growth rates, crystallographic orientation, and polytype selection in both films and nanowires during epitaxy growth.

Achieving the predictive and rational design of III-V semiconductor growth processes, rather than relying on empirical control, demands a quantitative framework that links atomic-scale events to macroscopic crystal growth behavior. For years, significant effort has been made to link growth parameters to morphology, composition, and crystal structure. Early analytical and numerical continuum frameworks~\cite{glas2013predictive,roper2007steady,tersoff2015stable,martensson2019simulation} established the growth rate equations that incorporate temperature, fluxes, droplet size, and contact angles, based on fitted parameters from experiments. On the other hand, first-principles \ac{MD} and \ac{MLIP}-based \ac{MD} simulations have revealed interface structures and ordering~\cite{sasaki2024temperature,zamani20203d}. However, these highly accurate methods are limited by time and length scale constraints, struggling to model realistic interfacial environments~\cite{magnussen2019atomicscale}. Beyond these, larger scale atomistic \ac{MD} simulations~\cite{soaresoliveira2020interatomic,oliveira2021role,chen2024diffusionlimited,wang2023analysis,liang2020investigation} have captured nucleation processes. Yet, standard \ac{MD} simulations are also limited by the time-scale issue, preventing the simulation of realistic growth predictions. Similarly, although \ac{KMC} simulations~\cite{spirina2024monte,chugh2017lattice,erwin2020atomic,kaufmann2016critical,su2023microscopic,reyes2013unified} have shown promise in predicting epitaxy growth rates and morphology, current models are often limited to \ac{VS} systems or lack parameterization from accurate energetics of solid-melt interfaces. The pivotal advance will be the creation of a unified, multi-scale framework that links realistic atomistic energetics to macroscopic growth, finally providing the predictive power to engineer next-generation semiconductor materials with precisely tailored functionalities on demand.

In this work, we bridge this critical gap by developing and experimentally validating a multiscale framework that directly links fundamental atomic interactions to macroscopic growth kinetics. Our approach addresses longstanding questions about the role of liquid ordering on crystal growth at different directions and polarity~\cite{guniat2019vapor}. We begin by applying \ac{MD} simulations based on \ac{MLIP}~\cite{takamoto2022universala} to confirm that multiple layers of Ga atoms at the GaN-Ga melt interface form structurally ordered and electronically charged configurations. Then the metadynamics simulations~\cite{barducci2008welltempered,bonomi2019promoting,tribello2014plumed} reveal the complete \acp{FEL} for an N adatom adsorption from Ga melt and its migrations at the GaN-Ga melt interface, strongly related to the ordered structures of liquid Ga layers. These calculations uncover a "Ga lubrication" mechanism, where the liquid Ga at the interface dramatically lowers the migration energy barrier for N adatoms compared to a clean surface. This provides a mechanistic explanation for the catalytic role of Ga droplets~\cite{gruart2019role,songmuang2010identification,zheng2008kinetic}. These atomistic-level insights provide the key to unlocking predictive control. By integrating these newly quantified energy barriers into \ac{KMC} simulations, we show that GaN growth under typical \ac{MBE} conditions operates in a diffusion-controlled, layer-by-layer regime. We then embed these physics into a continuous model based on the free energies and concentration gradients to make quantitative predictions of facet-dependent growth rates. Crucially, we validate this entire framework through \ac{MBE} growth experiments, where our predicted rates are confirmed by \ac{HRTEM} analysis. This work, therefore, establishes a governing principle: for systems with low solubility of key elements, like N in liquid Ga, the free energy profile for adatom adsorption and migration at the solid-melt interface is the decisive factor determining the crystal's growth kinetics and rate.


\section*{Results}
\begin{figure}[!ht]
    \centering
    \includegraphics[width=0.8\textwidth]{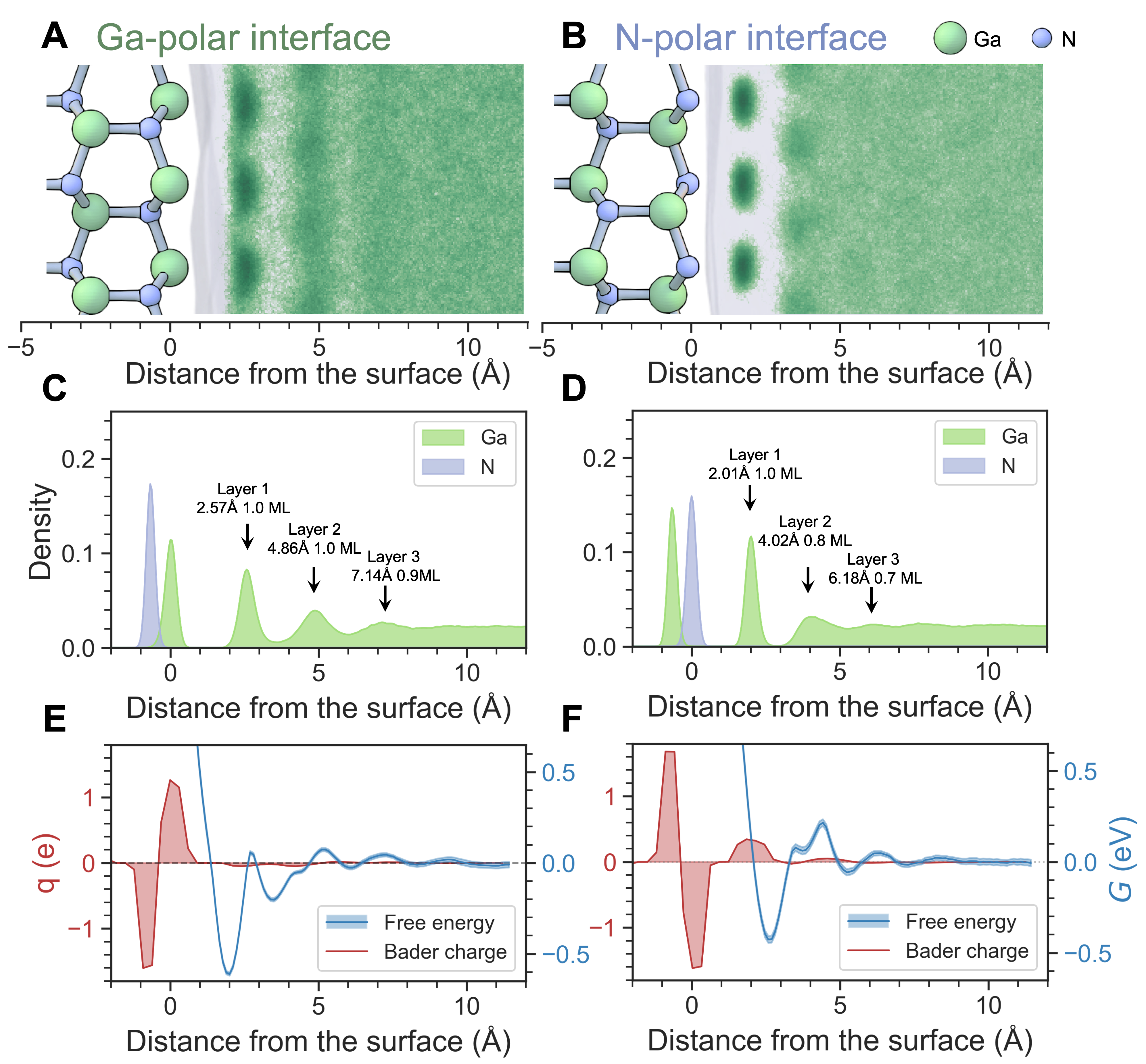}
    \caption{\textbf{Atomic structure, atomic density, charge distribution, and N-adatom energetics at Ga(\textit{l})-GaN(\textit{s}) interfaces.} (A)-(B): Atomistic configurations of the Ga-polar interface (A) and N-polar interfaces (B). Ga and N atoms are depicted in green and blue-gray, respectively. The liquid side is visualized by Ga atomic number density, computed from time-averaged histograms of Ga positions over long molecular-dynamics trajectories. This reveals the extent of ordering at the liquid Ga/GaN interface. (C)-(D): Averaged number density of Ga and N as a function of vertical distance from the surface of the Ga-polar (C) interface and N-polar interface (D), with distinct atomic layering annotated. (E)-(F): Bader charges (predicted from MLIPs) at the interface (red, left axis) and free energy changes of a single N adatom (blue, right axis) versus vertical position along the Ga-polar $\langle0001\rangle$ direction (E) and N-polar $\langle000\bar{1}\rangle$ direction (F).}
    \label{fig:order}
\end{figure}

\paragraph*{Solid-melt interface simulations} We performed \ac{MD} simulations to investigate interface-induced atomic ordering in liquid Ga at the GaN(\textit{s})-Ga(\textit{l}) interface (In this paper, \textit{l} and \textit{s} denote the liquid and solid states, respectively), employing a simulation geometry closely resembling the conditions found in laboratory experiments of GaN growth by self-catalyzed \ac{MBE}. An equilibrium GaN(\textit{s})-Ga(\textit{l}) interface was established, and long-duration equilibrium simulations were performed. This simulation setup enables an accurate determination of the average atomic density of Ga within the liquid phase. Fig.~\ref{fig:order} presents both the linear atomic density profiles projected across the simulation cell and the averaged atomic density of liquid Ga for Ga-polar and N-polar interfaces at 1000~K. The liquid Ga density (Figs.~\ref{fig:order} A-D) exhibits clear peaks in the direction normal to the interface, over several atomic layers. This interface-induced ordering of liquids is a well-known phenomenon~\cite{kaplan2006structural,freitas2020uncovering}, and we found that the extent and range of this ordering differ between the two polarities. Although the simulation geometry under \ac{MBE} conditions differs from the interfaces typically obtained under high-pressure bulk growth methods~\cite{dejong2020complex}, the observed ordering closely aligns with experimental observations from \ac{STM}~\cite{northrup2000structure}, in-situ X-ray \ac{CTR} scattering, and first-principles \ac{MD} simulations~\cite{sasaki2024temperature} for GaN. Similar ordering was also reported for other III-V systems during the gold-catalyzed or self-catalyzed nanowire growth at the solid-melt interfaces, including InP(\textit{s})-InAu(\textit{l})~\cite{algra2011formation}, GaAs(\textit{s})-Ga(\textit{l})~\cite{zamani20203d}, and GaAs(\textit{s})-GaAu(\textit{l})~\cite{oliveira2021role}.

The ordering observed at the interface originates primarily from interactions between Ga atoms in the liquid phase and the surface atoms of the GaN solid. On the Ga-polar interface (Fig.~\ref{fig:order}A), liquid Ga atoms bond with solid Ga atoms at the interface, forming clearly defined atomic layers. Specifically, `inner' Ga layers closer to the interface (Layer~1 and Layer~2 in Fig.~\ref{fig:order}C), comprise approximately one monolayer each, bonded to the underlying Ga atoms in the solid GaN, forming stable Ga-Ga pairs. Additional `outer' layers (Layers~3 in Fig.~\ref{fig:order}C) also emerge but exhibit weaker bonding, resulting in more fluid-like structures with reduced density. At the N-polar interface (Fig.~\ref{fig:order}B), the ordering is more short-ranged. Only the primary Ga layer directly bonded to the surface N atoms (Layer~1 in Fig.~\ref{fig:order}D) is well defined. The secondary Ga monolayer (Layer~2 in Fig.~\ref{fig:order}D) is present but laterally disordered and mobile, as indicated by lower density. This suggests weaker binding compared to the Ga-polar interface. Still, in both cases, more oscillations of the number density can be observed up to 1~nm from the surface. This simulated behavior aligns closely with experimental measurements~\cite{sasaki2024temperature}, underscoring the reliability and fidelity of our \ac{MD} simulations for GaN growth mechanisms.

Charge redistribution is much more localized than the ordering of the Ga liquid layers. Red curves in Figs.~\ref{fig:order}E and F show \ac{MLIP}-predicted Bader charge profiles, $q$, at the interfaces. The profile is computed by averaging per-atom charges with spatial binning over long-time equilibrium simulations. For both terminations, the dominant charge-transfer signal is confined to the primary, and at most the secondary, interfacial layer. On the Ga-polar interface, as the solid surface is positively charged Ga, liquid Ga presents slight negative charges, whereas on the N-polar interface, N termination induces positive charges for the adjacent Ga layer. By the third layer, the mean charge is indistinguishable from zero, even though number-density oscillations persist several angstroms into the liquid. This contrast indicates that at the  GaN(\textit{s})-Ga(\textit{l}) interfaces, electronic redistribution is governed by immediate interfacial bonding, while the atomic density peaks in liquid mainly reflect steric packing of finite-size Ga atoms driven by the bonding of the substrate~\cite{yan2020orientation}. 
 
\paragraph*{Free energy barriers for N adatom at the interfaces}
These ordered Ga layers significantly influence the growth dynamics at the liquid-solid interface, particularly affecting how the incoming N adatoms interact and incorporate into the solid lattice structure. Our MD simulations suggest that, during the formation of a new GaN layer, the ordered Ga layers and the related charge states can strongly affect the energetics of N adatom adsorption and migration to the energetically preferred adsorption sites. To quantify these energetics, we calculated detailed \ac{FEL} of an N adatom approaching both Ga-polar and N-polar interfaces at 1000~K (see Materials and Methods section for simulation details). 

\begin{figure}[!ht]
    \centering
    \includegraphics[width=0.8\textwidth]{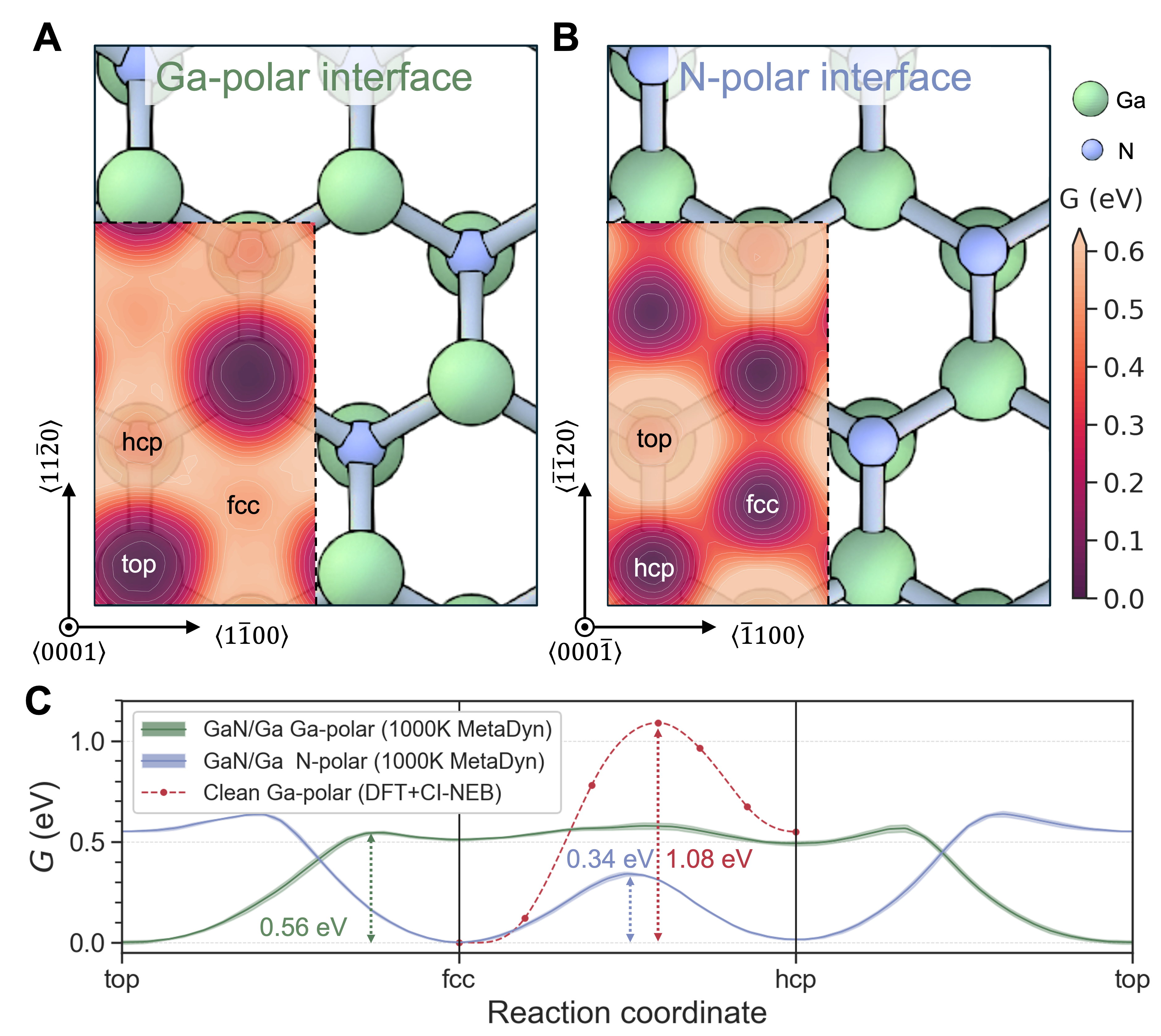}
    \caption{\textbf{Free energy profiles of an N adatom at Ga(\textit{l})-GaN(\textit{s}) interfaces.} (A)-(B): Atomic structures with contour maps of free energy landscapes for the Ga-polar (A) and N-polar (B) interfaces, superimposed on the atomistic structures with energy minima (adsorption sites labeled as top, hcp, and fcc). The most stable adsorption positions correspond to zero energy. (C): Comparative free energy profiles along the reaction coordinate, highlighting differences in energy barriers between Ga-polar and N-polar interfaces, and comparing with simulation results obtained from a clean Ga-polar surface.}
    \label{fig:energies}
\end{figure}

Blue curves in Figs.~\ref{fig:order}E and F present the free energy profiles for an N adatom approaching both Ga-polar and N-polar interfaces as a function of vertical distance from the surface, clearly illustrating how interface polarity impacts the energetic landscape for N incorporation. Each curve features distinct local minima and maxima, which align with the atomic density oscillations in Figs.~\ref{fig:order} A-D. For the Ga-polar interface, the most energetically favorable adsorption position (global minimum) occurs at approximately 2.0~Å from the interface, corresponding to a highly stable adsorption site immediately below Ga Layer~1. The secondary minimum is observed around 3.4~Å, corresponding to an intermediate position between Layers~1 and Layer~2, as shown in Fig.~\ref{fig:order}C. The intervening maxima coincide with liquid-layer density peaks, where the N adatom traverses high-density Ga liquid. Similarly, for the N-polar interface, the most favorable site lies at roughly 2.7~Å, situated between Layers~1 and Layer~2 as depicted in Fig.~\ref{fig:order}D, and the energy maxima coincide with the liquid-layer density peak of Layer 2. These energy profiles highlight the critical role of ordered liquid Ga layers in mediating N adatom adsorption, thereby significantly influencing the overall GaN growth kinetics, as we will discuss in the following section.

Figs.~\ref{fig:energies}A and B present free energy landscape contours as an N adatom migrates laterally at the GaN(\textit{s})-Ga(\textit{l}) interface. At the Ga-polar interface (Fig.~\ref{fig:energies}A), the most energetically favorable adsorption positions occur at `top' sites directly above surface Ga atoms. In contrast, at the N-polar interface (Fig.~\ref{fig:energies}B), these `top' sites are above surface N atoms and typically occupied by Ga atoms from the liquid Layer 1 (Fig.~\ref{fig:order}C). Therefore, threefold hollow sites, `fcc' and `hcp',  become the preferred adsorption sites for N adatoms at the N-polar interface. At these hollow sites, N adatoms effectively form bonds with surrounding liquid Ga atoms, which are themselves at `top' sites and bonded to the surface N atoms. The free energy barriers associated with migration among these lateral adsorption sites are shown explicitly in Fig.~\ref{fig:energies}C. On the Ga-polar interface, lateral migration of the N adatom occurs among adjacent `top' sites, involving passage through higher-energy intermediate positions at either `hcp' or `fcc' sites. This process exhibits an energy barrier of approximately 0.56~eV. By contrast, the N-polar interface enables the N adatom to migrate more directly between stable local minima at the `hcp' and `fcc' sites, with a significantly lower migration barrier of about 0.34~eV. Both barrier heights are relatively low compared to the thermal energy available at the typical \ac{MBE} growth temperature of 1000~K where $k_BT\approx0.086$~eV. Thus, N adatom diffusion across the interface is energetically feasible under typical growth conditions, indicating a smooth surface morphology during GaN \ac{MBE} growth.

The lower migration barriers at the interface can be attributed primarily to the presence of liquid Ga atoms, which form Ga-N bonds that facilitate the lateral migration of N adatoms. This corresponds to excess-group-III environments for self-catalyzed epitaxy growth~\cite{calleja2000luminescence,li2006directiondependent,bertness2008mechanism,guo2010catalystfree,moustakas2011invited}. Without liquid Ga, the migration of N adatoms on a clean GaN surface is significantly more difficult. As shown in Fig.~\ref{fig:energies}C, the \ac{DFT} + \ac{CI-NEB} calculated migration barrier on a clean Ga-polar surface is about $\sim$1.1eV, substantially higher than at the liquid-covered interfaces. On this clean surface, N adatoms strongly prefer the `fcc' sites, where they form three stable bonds with surface Ga atoms, consistent with previous theoretical findings~\cite{takeuchi2005adsorption,chugh2017lattice}. Without liquid Ga, N adatoms become trapped at energetically favorable yet incorrect `fcc' positions, impeding their migration to the optimal growth `top' sites. For clean surface calculations, we also applied \ac{CI-NEB} with \ac{MLIP} and metadynamics simulations to verify the reliability and accuracy of our \ac{MLIP}-based calculations. (see Supplementary Text S1 for more details).

\begin{figure}[!ht]
    \centering
    \includegraphics[width=0.8\textwidth]{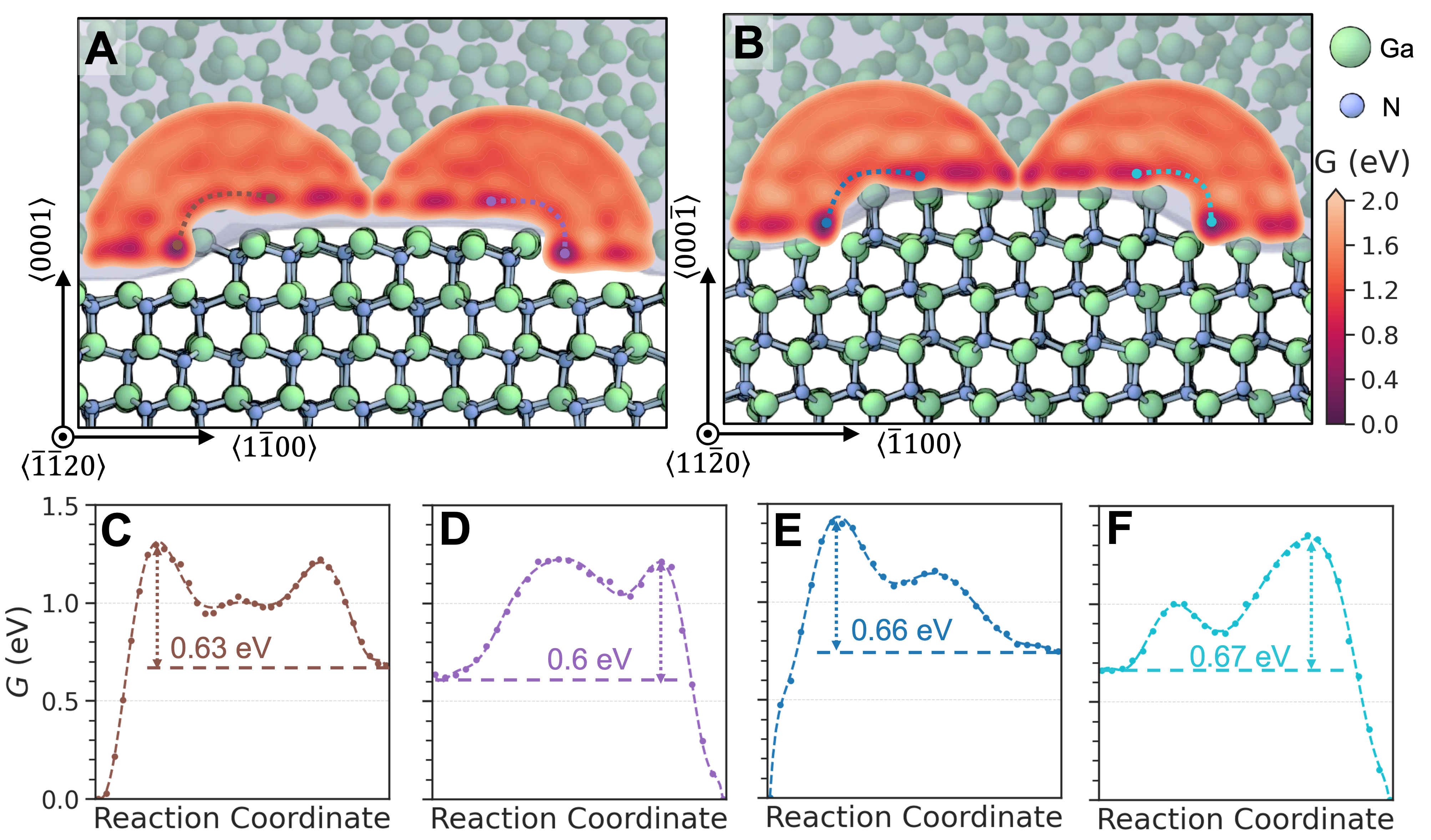}
    \caption{\textbf{Free energy landscape and step-edge barriers for N adatom migration on Ga(\textit{l})-GaN(\textit{s}) interfaces with step-terrace structures.} (A)-(B): Free energy contours illustrating the energetics of N adatoms interacting with  $\langle1\bar{1}00\rangle$ steps at the Ga-polar (A) and N-polar (B) interfaces. The atomistic models visualize the corresponding interfacial structures. (C)-(F): Free energy profiles along the reaction coordinate for each step-edge, highlighting the barriers for adatom migration across terrace steps.
    }
    \label{fig:step}
\end{figure}

In addition to examining adsorption and migration free energies on smooth interfaces, we employed metadynamics simulations to investigate the step-edge migration barriers at Ga(\textit{l})-GaN(\textit{s}) interfaces featuring step-terrace structures. Figs.~\ref{fig:step}A and B illustrate the atomic configurations of the $\langle1\bar{1}00\rangle$ structures, along with the corresponding free energy contours that reveal the interaction landscape for N adatoms near the steps. An N adatom approaching from the lower terrace can bond directly to the step-edge site, which is energetically favorable due to its lower free energy and greater coordination with neighboring atoms. In contrast, N adatoms on the terrace away from the step exhibit fewer nearest-neighbor bonding and exhibit higher energy states. When an atom approaches the step from the upper terrace, it encounters a significant energy barrier ($>$0.6~eV, Figs.~\ref{fig:step}C-F), which is higher than the diffusion barrier calculated on the flat interface (0.34~eV and 0.65 eV, Fig.~\ref{fig:energies}C). These additional energy barriers, known as \ac{ES} barriers~\cite{ehrlich1966atomic,schwoebel1966step}, can hinder adatom descent across step edges. When they are sufficiently high, they lead to adatom accumulation on upper terraces, inhibiting smooth layer-by-layer growth and instead promoting the formation of rough surface morphologies such as islands or mounds during GaN growth.

\begin{figure}[!ht]
    \centering
    \includegraphics[width=1.0\textwidth]{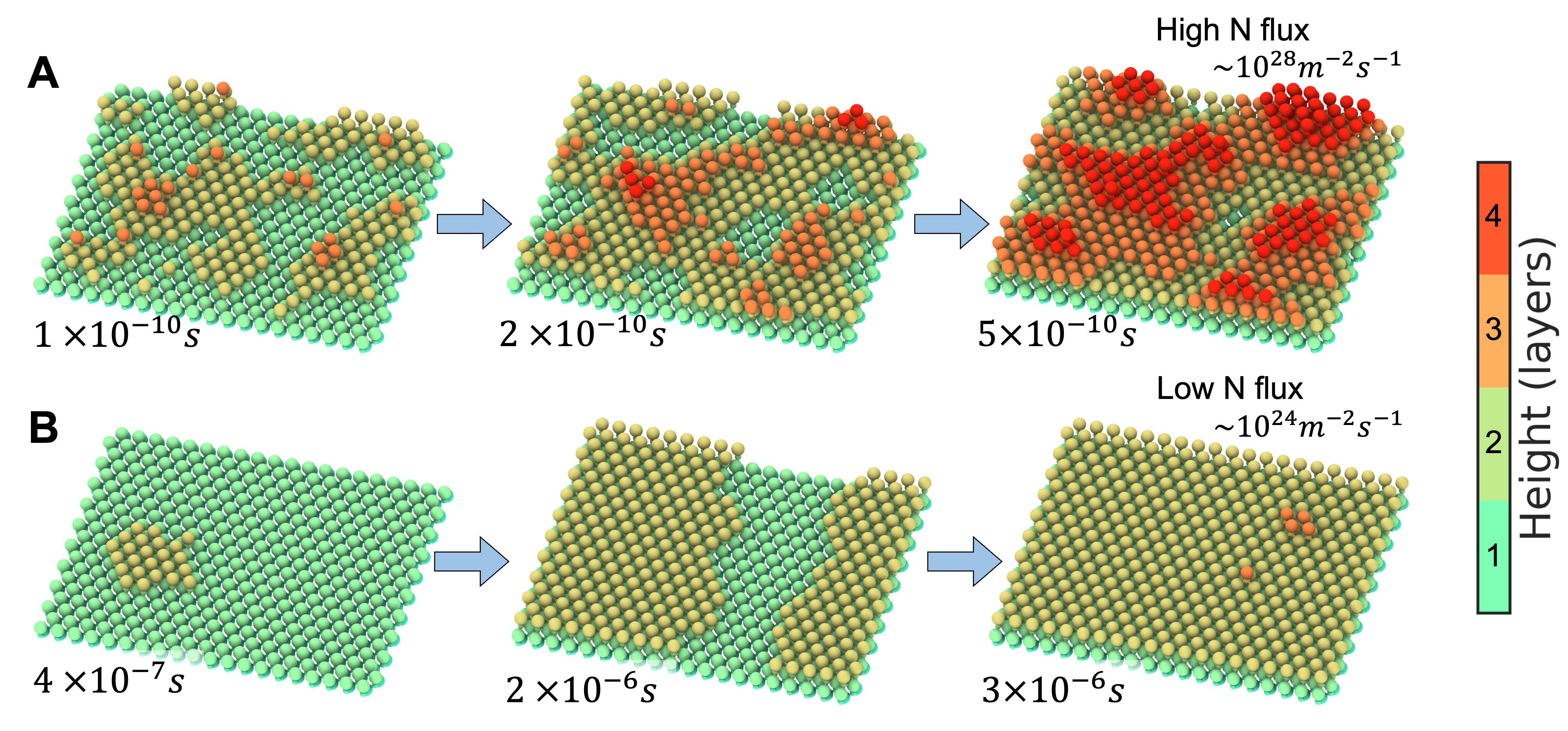}
    \caption{\textbf{Perspective views of GaN crystal growth at the Ga(\textit{l})-GaN(\textit{s}) interface from KMC simulations.} (A): Under a high N flux ($10^{28}$~m$^{-2}$s$^{-1}$), growth occurs via a joint-island mechanism, resulting in the formation of multiple nucleation islands that grow vertically. (B): At a lower N flux ($10^{24}$~m$^{-2}$s$^{-1}$, note this is still orders of magnitude higher compared to typical experimental conditions), a layer-by-layer growth mode dominates, characterized by the lateral spreading and completion of one atomic layer before subsequent layers form. The color scale represents the height of the grown layers.}
    \label{fig:kmc}
\end{figure}

\paragraph*{Growth mode analysis via kinetic Monte Carlo simulations}
To bridge the atomistic energetics to macroscopic growth phenomena, we applied \ac{KMC} simulations on a \ac{WZ} ($0001$) plane, parameterized directly by the migration and step-descent barriers obtained above (see Materials and Methods section for simulation details). We examined two regimes of N flux from the Ga melt to the GaN surface. Fig.~\ref{fig:kmc} presents the resulting morphologies and highlights a clear N flux-dependent difference in growth mode. As shown in Fig.~\ref{fig:kmc}A, with high nitrogen flux of $10^{28}$~m$^{-2}$s$^{-1}$, adatoms reach the GaN surface rapidly. Many nuclei appear across the surface and grow vertically, producing a rough, joint-island texture with many vertical mounds. Since adatoms reach the interfaces much faster than they can laterally diffuse and step down the upper terraces, N incorporation is limited by the interface-controlled regime. On the other hand, with a lower nitrogen flux of $10^{24}$~m$^{-2}$s$^{-1}$, adatoms arrive more slowly, allowing them to explore the surface and settle into the lowest-energy sites before another layer is supplied. Therefore, this diffusion-controlled growth advances by orderly completion of each monolayer.

In actual \ac{MBE} growth conditions, the GaN growth rate is on a much smaller scale of $\sim$0.01~nm/s~\cite{riechert1996mbe,wang2004molecular}, which implies that the effective N incorporation rate is on the scale of $\sim10^{18}~\text{m}^{-2}\text{s}^{-1}$. Though the incoming active N flux must be higher because many N species may bounce off or diffuse away at the surface, and only a fraction of the incoming atoms are captured at sites on the GaN surface. As shown in the Figs.~\ref{fig:order}E and F, there are barriers of $\sim$0.1 to 0.2~eV at the WZ polar basal planes for incoming N adatoms as they approach the surface. At 1000~K, this corresponds to an active N flux of $\sim$ $(3$ to $10)\times 10^{18}~\text{m}^{-2}\text{s}^{-1}$.
These fluxes remain several orders of magnitude lower than the `low N flux' used in our \ac{KMC} simulations. Thus, under practical \ac{MBE} conditions, we expect the same diffusion-limited step-flow regime, generating smooth terraces, efficient step-edge capture, and minimal interfacial defect formation. 

\paragraph*{Diffusion-controlled crystal growth model}
Thus, a mathematical model for diffusion-controlled crystal growth, derived from a modified version of Fick's first law and taking into account the chemical potential gradient at the interfaces, was also developed to describe the steady-state flux of N adatoms as well as the NW growth rates.

To derive the steady-state flux, $J_{\text{N}}$, of N adatom diffusing and adsorbing at the GaN(\textit{s})-Ga(\textit{l}) interface, we start from a master equation of the Smoluchowski equation, a modified form of Fick's first law with an external potential~\cite{balluffi2005kinetics}:
\begin{equation}
    J_{\text{N}} = -D \frac{dc\left(z\right)}{dz} - \frac{Dc\left(z\right)}{k_B T} \frac{dG\left(z\right)}{dz}
    \label{eq:master}
\end{equation}
where $D$ is the diffusivity of N in liquid Ga, $c\left(z\right)$ and $G\left(z\right)$ denote the concentration and chemical potential of N atom at the distance $z$ from the interface. The first term in the equation is the contribution of the concentration gradient, and the second term incorporates the influence of chemical potential gradients on atomic diffusion. By applying appropriate boundary conditions of fixed chemical potential and fixed concentration (see Supplementary Text S5 for more details), we obtain the solution to Eq.~\ref{eq:master} for steady-state N flux:
\begin{equation}
    J_{\text{N}} = D \left( \frac{ c\left(0\right) \exp \left( \frac{G\left(0\right)}{k_B T} \right) - c\left(L\right) \exp \left( \frac{G\left(L\right)}{k_B T} \right)}{\int_0^{L} \exp \left( \frac{G(z)}{k_B T} \right) dz} \right)
    \label{eq:flux}
\end{equation}
In this equation, $G(z)$ can be determined through metadynamics simulations, as illustrated in Figs.~\ref{fig:order}E and F. The position $z=L$ is located sufficiently far from the interface within the liquid, and the corresponding concentration $c\left(L\right)$ represents the equilibrium solubility limit of N in Ga liquid. At the interface (position $z=0$), the concentration $c\left(0\right)$ represents the concentration of N in the liquid phase right at the interface. Due to the rapid interfacial reaction as indicated by our \ac{KMC} simulations (Fig.~\ref{fig:kmc}B), we assume that N adatoms reaching the interface immediately migrate into energetically favorable sites and incorporate into the solid GaN lattice without accumulating in the liquid phase. This rapid interfacial reaction implies that the interfacial liquid concentration is effectively zero. Thus we set $c\left(0\right) = 0$.

With Eq.~\ref{eq:flux}, we can now calculate the facet-normal growth rate directly. This equation is independent of the choice of the free energy reference. Thus, by setting the liquid reference so that $G(L)=0$, the flux reduces to a simpler form
$J_{\text{N}}=Dc(L)\big/\int_0^{L}\exp\left[G(z)/k_BT\right]dz$. In this equation, the numerator represents the bulk supply capacity of nitrogen to the interface, which is the same for all different facets. The denominator acts as an interfacial resistance that is entirely determined by the metadynamics free-energy profile of the chosen facet. All quantities in Eq.~\ref{eq:flux} can be obtained directly from atomistic simulations without fitted parameters. The interfacial free‐energy profile $G(z)$ comes from well-tempered metadynamics (Figs.~\ref{fig:order}E and F). The diffusivity $D(T)$ is calculated from the Einstein relation based on MD simulation results (See Supplementary Text S3). Concentration of N in the Ga liquid $c(L)$ is converted from the mole fraction of dissolved N $x_\text{N}$, which is estimated by the solvation free energy $\Delta G_{\text{solv}}^{\text{N}}$ (See Supplementary Text S4). After obtaining flux, the growth rate of a facet $f$ is $v_f$ defined as the following:
\begin{equation}
v_f=\frac{J_{\text{N}}}{n_s^{(f)}}h^{(f)},
\end{equation}
with $n_s^{(f)}$ the surface site density and $h^{(f)}$ the step height for adding one growth monolayer on the facet $f$. In the following, we compare the results of $v_f$ for different GaN facets in both \ac{WZ} and \ac{ZB} phases and compare them directly with experimental observations.

\begin{figure}[!ht]
    \centering
    \includegraphics[width=0.6\textwidth]{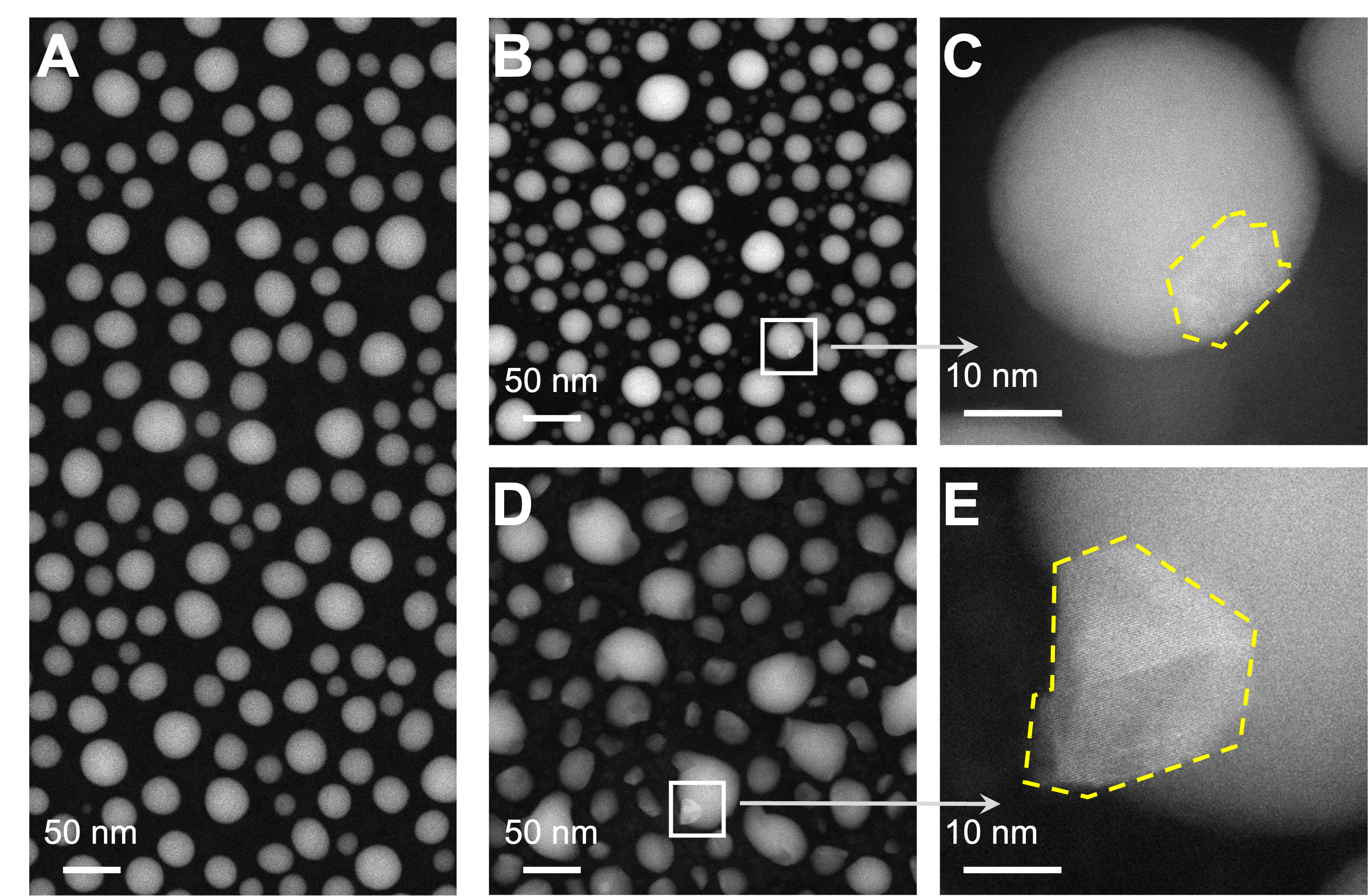}
    \caption{\textbf{Nanostructures of Ga droplets and partially nitridated GaN/Ga particles.} (A)-(E): HAADF-STEM images of (A) Ga droplets after Ga pre-deposition, (B-C) particles after 1~min nitridation, and (D-E) particles after 5~min nitridation in different magnifications. Yellow dashed lines marked crystallized GaN regions within particles. White solid boxes highlight the magnified regions.}
    \label{fig:stem}
\end{figure}

\paragraph*{Growth of GaN particles through Ga-mediated MBE}
To quantify the GaN growth rate in the self-catalyzed \ac{VLS} method, we prepared partially nitridated Ga/GaN particles by \ac{MBE}. We first investigate the morphology of Ga droplets and partially nitridated GaN/Ga particles under 1~min and 5~min exposure of N plasma in \ac{MBE}. \Ac{HAADF-STEM} images are presented in Fig.~\ref{fig:stem}. Fig.~\ref{fig:stem}A reveals a uniform ensemble of smooth, hemispherical Ga droplets on amorphous SiN$_x$ prior to nitridation, with diameters on the order of tens of nanometers. After a short 1~min exposure to the N$_2$ plasma at 650$^{\circ}$C and $5\times10^{-7}$~Torr, the particles largely retain their circular profiles (Fig.~\ref{fig:stem}B). Only a small fraction of Ga droplets crystallize into GaN, as indicated by the yellow dashed line marking the crystallized region in Fig.~\ref{fig:stem}C.  Extending the nitridation to 5~min resulted in significant morphological changes, as shown in Fig.~\ref{fig:stem}D. The particles transitioned from a circular to a faceted shape due to the extended plasma exposure. Fig.~\ref{fig:stem}E reveals the formation of facets within a Ga droplet. 

We then analyzed the crystallographic information and GaN radius of over 200 particles from the 1~min and 5~min samples. We observed different polytypes and crystal planes parallel to the substrate surface, which are perpendicular to the electron beam in \ac{TEM}. More than five crystal planes were observed in 1 min and 5 min GaN/Ga particles including \ac{WZ}$(0001)$, \ac{WZ}$(1\bar101)$, \ac{WZ}$(1\bar100)$, \ac{WZ}$(11\bar23)$, \ac{ZB}$(001)$, and \ac{ZB}$(110)$ (See Supplementary Text S6 for crystallographic statistics). 

\begin{figure}[!htbp]
    \centering
    \includegraphics[width=0.8\textwidth]{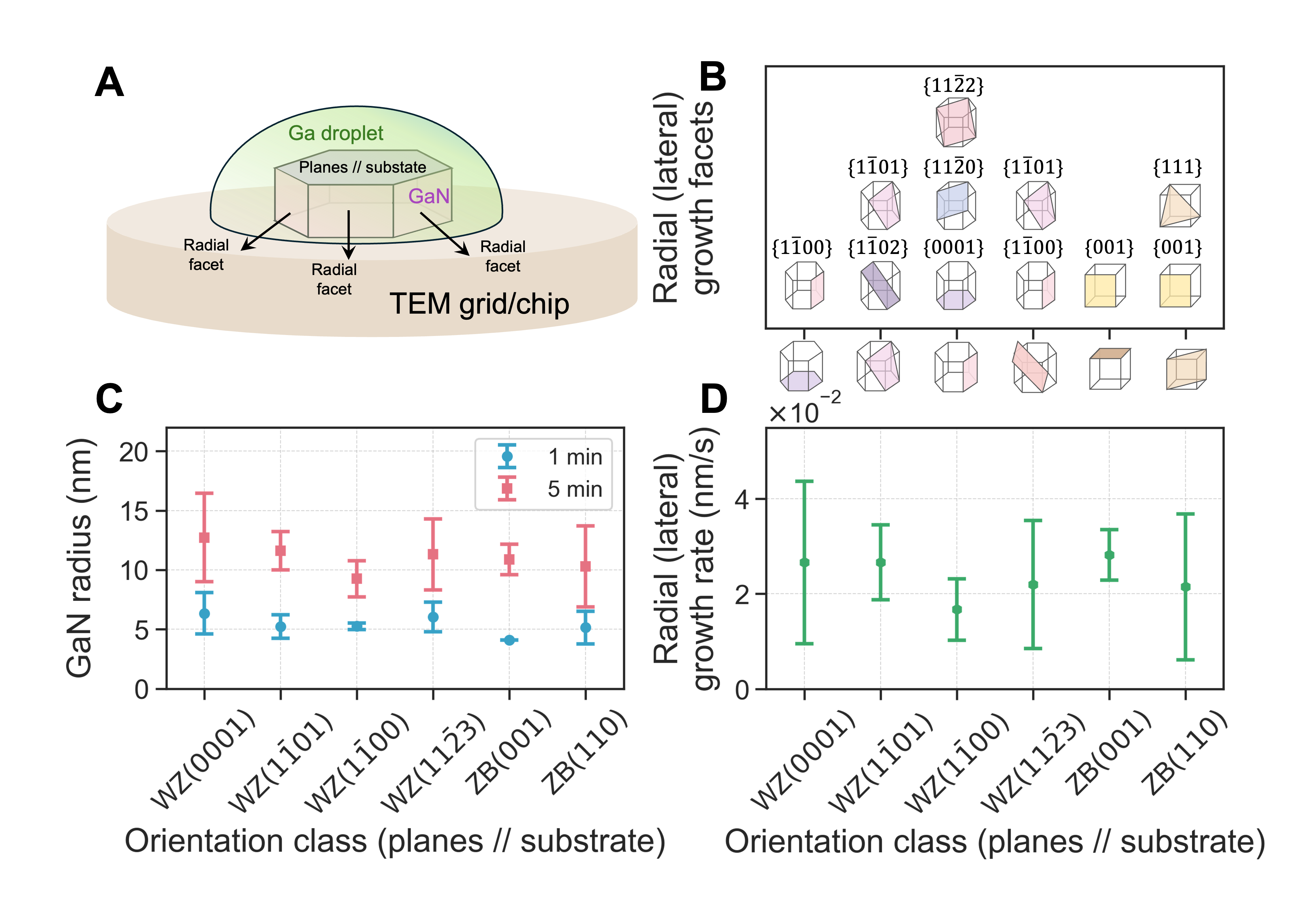}
    \caption{\textbf{Summary of GaN growth rates at the interfaces from TEM results.} (A): Schematic of a GaN particle in a Ga droplet on the TEM chip, highlighting planes parallel to the substrate and the external side facets that were measured by TEM.(B): Summary of observed external facets identified for each orientation class from TEM. (C): Mean GaN radius measured after 1 and 5 min nitridation for particles grouped by the plane parallel to the substrate; points are means and error bars reflect standard deviations. (D): Facet-dependent experimental growth rates inferred from the radii in (C).}
    \label{fig:exp_growth_rate}
\end{figure}

\paragraph{GaN growth rates at different facets}
In addition to indexing the crystal planes parallel to the substrate, we identified the external growth facets that define each particle’s morphology (see Supplementary Text S6 for more details). The facets represent the outer crystal planes that define the external shape of the particles, as shown in the schematic in Fig.~\ref{fig:exp_growth_rate}A. For example, for a \ac{WZ} GaN particle with a $(0001)$ basal plane, prismatic m-planes $(1\bar100)$ frequently bound the sidewalls (hexagonal morphology); likewise, \ac{ZB} particles with $(001)$ top facet are commonly terminated by $(001)$ side facets (cubic morphology). A summary of the external facets identified for each particle orientation is provided in Fig.~\ref{fig:exp_growth_rate}B.

We used these assignments to extract orientation-dependent growth rates by measuring the average GaN radius after 1 and 5~min of nitridation. The radii and the corresponding rates are plotted in Figs.~\ref{fig:exp_growth_rate}C and D, respectively. Across all orientations, the mean radius spans $\sim$3-7~nm at 1~min and $\sim$10-15~nm at 5~min, consistent with the observations in Figs.~\ref{fig:stem}C and E.  Importantly, each value represents an ensemble average over the different facets present on particles in that orientation class as shown in \ref{fig:stem}A. Hence, the results are grouped by the directions of planes parallel to the substrate surface. The error bars reflect the inter-particle spread and facet-to-facet variability.

\begin{figure}[!htbp]
    \centering
    \includegraphics[width=0.8\textwidth]{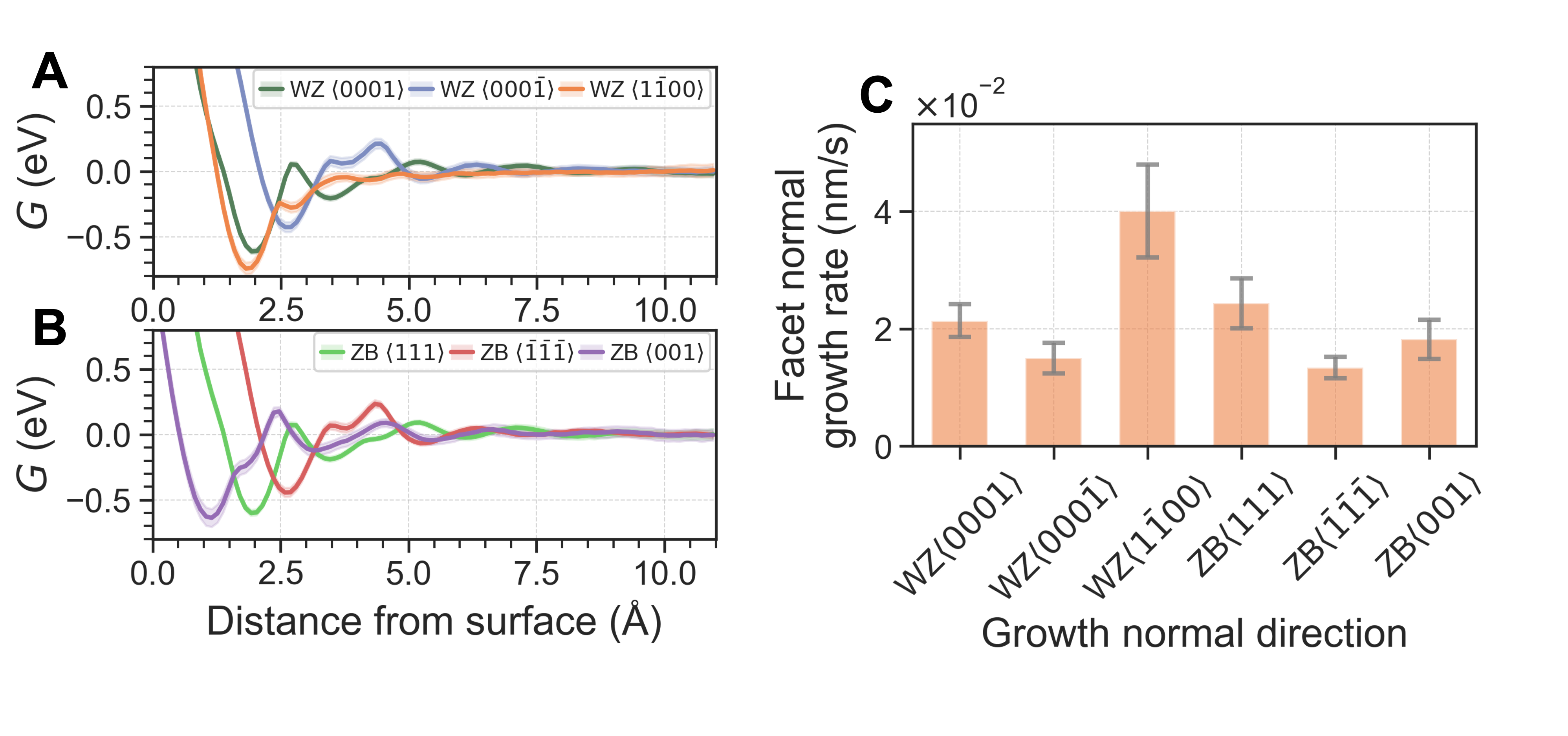}
    \caption{\textbf{Summary of GaN growth rates at the interfaces from growth model.} (A)-(B): One-dimensional free-energy profiles $G(z)$ for N adatom adsorption normal to the interface, obtained by well-tempered metadynamics for representative WZ facets (A) and ZB facets (B). (C): Predicted steady-state facet growth rates from the parameter-free diffusion model (Eq.~\ref{eq:flux}) using the free-energy profiles $G(z)$, N diffusivity $D(T)$, and N solubility limit $c(L)$. Error bars are estimated based on the error shade from the free energy curves. The calculated rates (0.01-0.04 nm/s) reproduce the facet trends observed experimentally in Fig.~\ref{fig:exp_growth_rate}D.}
    \label{fig:simu_growth_rate}
\end{figure}

For comparison, we computed facet-resolved adsorption free-energy profiles $G(z)$ at the GaN(\textit{s})-Ga(\textit{l}) interface via metadynamics (Fig.~\ref{fig:simu_growth_rate}A and B for \ac{WZ} and ZB, respectively) and brought $G(z)$ into Eq.~\ref{eq:flux}. The resulting diffusion-controlled model predicts facet-specific growth rates in the range, varying from 0.01 nm/s to 0.04 nm/s on different facets as shown in Fig.~\ref{fig:simu_growth_rate}C. Notably, the rates are directly calculated from the solid-liquid interfaces; hence, they represent the rates of planes perpendicular to the growth directions. The experimental bars in Fig.~\ref{fig:exp_growth_rate}D are grouped by the plane parallel to the substrate (each group contains an ensemble of external facets as shown in Fig.~\ref{fig:exp_growth_rate}B) and several facets visible in TEM, in particular WZ ${11\bar20}$ and the semipolars ${10\bar11}$ and ${11\bar22}$ were not sampled. Despite those issues, some general trends are observed when comparing with the facet-resolved predictions in Fig.~\ref{fig:simu_growth_rate}C. For example, within \ac{WZ}, prismatic m-plane $(1\bar{1}00)$ grows faster than basal planes, and the two polar basal planes are close in rate. This matches the higher radii and higher growth rates for particles whose sidewalls are dominated by m-planes (WZ$(0001)$ particles) relative to those bounded by basal facets (WZ$(1\bar{1}00)$ particles). Within \ac{ZB}, the model predicts the growth rate of Ga-polar plane $(111)$ is faster than N-polar plane $(\bar1\bar1\bar1)$, with $(001)$ intermediate. Experimentally, particles with $(001)$ parallel to the substrate only present ${001}$ sidewalls, and these groups display similar growth rates with less standard deviation than the ZB $(110)$ particles, which are bounded by ${001}$ and ${111}$. 

Together, the TEM statistics and facet-resolved free-energy calculations show that GaN growth under our MBE conditions is extremely slow due to diffusion-limited N transfer. This process is governed by the interfacial free-energy landscape, $G(z)$. Facets with deeper driving forces and smaller barriers exhibit larger radii and higher rates, and the parameter-free predictions from Eq.\ref{eq:flux} (Figs.~\ref{fig:simu_growth_rate}A and B) quantitatively capture this trend. This agreement validates our multiscale framework, which links atomistic energetics to macroscopic morphology, and provides a predictive basis for tuning facet evolution in self-catalyzed III-V growth.

\section*{Discussion}
These findings deepen our understanding of GaN growth in \ac{MBE} by mapping multiscale simulations to experimentally measured growth kinetics. First, equilibrium \ac{MD} reveals pronounced layering of liquid Ga in contact with GaN (Fig.~\ref{fig:order}), which reorganizes the local bonding environment experienced by incoming N adatoms. Second, well-tempered metadynamics quantifies adsorption (Figs.~\ref{fig:order} E and F), lateral migration (Fig.~\ref{fig:energies}, and step-edge barriers (Fig.~\ref{fig:step})on specific facets and polarities, uncovering a ``Ga lubrication'' mechanism whereby interfacial Ga layers substantially reduce N migration barriers compared with the clean surface. Third, passing these barriers into KMC simulations demonstrates a diffusion-controlled layer-by-layer growth mode for self-catalyzed GaN growth in \ac{MBE} conditions. Thus, the same free-energy inputs are applied in a parameter-free transport equation (Eq.~\ref{eq:flux}). Combined with the \ac{MD}-computed N diffusivity and free-energy-calculated N solubility in Ga, it predicts facet-resolved growth rates without any fitted parameters. The predicted rates and their facet dependence agree with the radii extracted from \ac{STEM}/\ac{HRTEM} statistics (Fig.~\ref{fig:exp_growth_rate}). Taken together, these findings provide a unified multiscale picture of GaN growth at metal/semiconductor interfaces.

This GaN growth picture contrasts sharply with other III-V systems such as InP~\cite{algra2011formation} or GaAs~\cite{reyes2013unified,oliveira2021role}. In such systems, P and As are orders of magnitude more soluble in metal liquid at growth temperatures than N in Ga~\cite{ansara1994binary}, allowing the droplet composition and supersaturation to be tuned by the V/III flux. Those changes in supersaturation and contact angle at the triple-phase line often control polytype selection and switching, as well as axial versus radial growth modes~\cite{glas2013predictive,jacobsson2016interface}. For GaN and other nitride systems, by contrast, the droplet remains overwhelmingly metal-rich and near zero N solubility~\cite {jones1984liquidus,onderka2002thermodynamics}. Polytype selection and facet evolution therefore, cannot be efficiently modulated through bulk chemistry; instead, the growth of nitride is governed by the diffusion and interface-introduced adsorption barriers encoded in $G(z)$.

Therefore, our workflow is greatly transferable. Every input in Eq.~\ref{eq:flux} is computed directly from atomistic simulations: $G(z)$ from metadynamics, $D(T)$ from equilibrium \ac{MD}, and $c(L)$ from the two-step N$_2$ solvation/dissociation free energy. As a result, the model is predictive rather than interpretive. All information can be recomputed for a different III-V/catalyst system (e.g., InN(\textit{s})-In(\textit{l}) or GaN(\textit{s})-GaAu(\textit{l})) under new temperatures and partial pressures, as no empirical fitting to growth data is required. More broadly, the multiscale workflow provides a transferable template for predictive crystal growth kinetics in other systems with low solubility of key elements.

Despite the advances enabled by our integrated multiscale framework, some limitations remain in the current implementation. First, impurities such as O or H, which are known to incorporate during nitride growth, were not explicitly considered. Even though the base pressure of \ac{MBE} can reach as low as 10$^{-10}$ Torr, trace residual impurities may still perturb interfacial energetics by modifying local bonding environments, competing for adsorption sites, or introducing trap states that alter diffusivity and incorporation kinetics. Second, it is important to recognize the limitations in the quality and transferability of the employed \acp{MLIP}. While the \ac{PFP} has demonstrated strong accuracy across a wide chemical space, any \ac{MLIP} is ultimately limited by its training set and may fail to capture rare configurations, high-energy defects, or plasma-excited species encountered in real \ac{MBE} conditions. Further benchmarking against first-principles calculations and more experimental measurements will therefore be critical to validate the robustness of these predictions. Third, our diffusion-controlled growth model (Eq.~\ref{eq:flux}) relies on simplifying assumptions, including steady state, a one-dimensional reaction coordinate $z$, constant diffusivity $D(T)$, and zero surface concentration $c(0)=0$ (no interfacial accumulation). It also neglects a lot of details at the interfaces, such as lateral heterogeneity, capillarity, curvature effects, and triple-phase-line kinetics. Finally, the experimental statistics average over multiple facets within each orientation class, which makes it difficult to isolate facet-specific growth rates. Future work combining \textit{in situ} microscopy, together with models that couple interfacial barriers to droplet dynamics, supersaturation, and impurity chemistry, will be essential to further refine and test this framework.

In conclusion, our results establish a direct, quantitative link between atomic-scale energetics at the solid-melt interface and experimentally observed growth rates and morphologies. The derived insights into GaN growth kinetics offer a predictive framework and provide actionable levers to steer semiconductor growth direction, polarization, morphology, polytype, and ultimately the electronic and optical performance of semiconductors such as the Ga-N system.

\section*{Materials and Methods}
\paragraph*{Molecular dynamics simulations}
We performed \ac{MD} simulations using \ac{PFP}~\cite{takamoto2022universala,takamoto2023universala} with \ac{LAMMPS}~\cite{plimpton1995fast,thompson2022lammps}. \ac{PFP} is a highly accurate universal neural network interatomic potential integrated within the Matlantis software package~\cite{matlantis}. The timestep was selected as 1~fs. First, an 8-layer, 768-atom orthorhombic supercell of \ac{WZ} GaN was equilibrated for 3~ps using a Langevin thermostat with a damping parameter of 0.1~ps and a Nosé-Hoover barostat with a damping parameter of 1~ps. The equilibrium lattice constant was then determined by averaging over a subsequent 2~ps run. Next, an interface model was constructed by placing the equilibrated GaN slab in contact with 500 liquid Ga atoms, which had been separately equilibrated for 5~ps using the same Langevin thermostat. This entire solid-melt interface system was equilibrated for 5~ps, followed by a 1~ns production run where configurations were recorded every 1~ps. The simulations were performed at 1000~K, a typical growth temperature for GaN in \ac{MBE}.
\paragraph{Metadynamics simulations}
We computed the free energy for N at the Ga(\textit{l})-GaN(\textit{s}) interface using well-tempered metadynamics~\cite{barducci2008welltempered} implemented via the PLUMED library~\cite{bonomi2019promoting,tribello2014plumed} coupled to \ac{LAMMPS}. Metadynamics simulations applied the same \ac{MLIP} and settings as our equilibrium MD (time step 1~fs, $T=1000$~K, Langevin thermostat with 0.1~ps damping). Several interface slabs are modeled, including Ga-polar \ac{WZ} $(0001)$ and \ac{ZB} $(111)$, N-polar \ac{WZ} $(000\bar1)$ and \ac{ZB} $(\bar1\bar1\bar1)$, non-polar \ac{WZ} $(1\bar100)$ and \ac{ZB} $(001)$. Each slab contained a solid GaN region of 768 atoms (512 atoms for \ac{ZB} $(001)$) and a liquid Ga region with 500 atoms. Then a single N atom was placed at the solid-melt interface. Each system was equilibrated for 5~ps at 1000~K prior to biasing.

We applied well-tempered metadynamics with an initial hill height of 0.01~eV, direction-dependent widths (see below), and a bias factor of 5. The simulation ran for 10 ns, during which Gaussian hills were deposited every 0.4 ps. To prevent the N atom from drifting deep into the bulk liquid or desorbing far from the interface, the system was constrained by soft harmonic walls placed sufficiently far from the reaction coordinate to not influence the calculated free energy profile.

To calculate the vertical adsorption free energies (Figs.~\ref{fig:order}E and F), we used a single a single \ac{CV} defined as the distance between the N atom and the instantaneous center of mass of the top GaN layer. For this \ac{CV}, we applied a Gaussian width of $0.20$~Å. To determine the profile and its statistical error, we used a block averaging method. This analysis is based on the Tiwary-Parrinello formalism~\cite{tiwary2015timeindependent}, where the trajectory is first reweighted before being divided into five blocks for error analysis. The final profile and confidence intervals were derived from the mean and standard error across these blocks. 

To calculate the lateral migration free energies (Fig.~\ref{fig:energies}), we performed two-dimensional metadynamics simulations using the in-plane coordinates $(x,y)$ of N atom within the surface unit cell as \acp{CV}. These correspond to the projections along the two in-plane lattice vectors of the given surface. Gaussian width is $0.20$~Å in each dimension. Periodic boundary conditions were applied to the \acp{CV} so that $(x,y)$ remained confined to the first unit cell, enabling more efficient sampling. The resulting two-dimensional free-energy surface was obtained by reweighting to this reference cell. Migration barriers between adjacent minima (e.g., `top'$\leftrightarrow$`top' on Ga-polar, `hcp'$\leftrightarrow$`fcc' on N-polar) were extracted as the minimum free-energy heights along the minimum-free-energy pathway.

To investigate diffusion over a step edge, we constructed supercells with steps along the $\langle1\bar{1}00\rangle$ direction. A nitrogen probe atom was initially placed on the upper terrace near the step. We then calculated the two-dimensional free-energy surface using well-tempered metadynamics with two \acp{CV}: the lateral $x$ and vertical $z$ distance of the N atom relative to the step edge. Gaussian hills with a width of $0.20$~Å were used for each \ac{CV}. A weak harmonic restraint was also applied to keep the N atom near the step. The final two-dimensional free-energy surface was obtained by reweighting the bias potential, and from this, we extracted the minimum free-energy pathway for the N atom crossing the step edge.

\paragraph{Kinetic Monte Carlo simulations}
We modeled the growth of GaN on the $(0001)$ interface using rejection-free \ac{KMC} simulations. Our model employed a three-dimensional \ac{WZ} crystal structure within an orthorhombic simulation cell, with periodic boundary conditions applied along the in-plane directions. The initial substrate consisted of two GaN bilayers on a $10\times20$ unit cell grid. Each layer consists of 400 N atoms and 400 Ga atoms.

The simulations assume Ga-rich conditions with the presence of Ga droplet on the surface. This implies that the availability of Ga is not a rate-limiting step, as Ga atoms can readily bond with N atoms at the growth interface. Therefore, our \ac{KMC} simulations focus on the kinetics of N. The elementary kinetic events included in the model are: (i) the hopping of an N adatom to a nearest-neighbor site on the same terrace; (ii) the crossing of an N adatom over a step edge to an adjacent upper or lower terrace, which includes an additional \ac{ES} energy penalty; and (iii) the adsorption of a new N adatom onto a randomly selected surface site.

The event rates ($r_\text{mig}$) for the first two migration events are calculated using the Arrhenius equation: 
\begin{equation}
    r_\text{mig} = \nu \exp(-E_a/k_B T).
\end{equation}
Here, the attempt frequency $\nu$ was set to be at a typical crystal frequency of $10^{13}$ Hz, the growth temperature $T$ was 1000~K, and $k_B$ is the Boltzmann constant. The free energy barriers of this event, $E_a$, were determined using the \ac{BEP} principle~\cite{bronsted1928acid,evans1935applications}, which establishes a linear relationship between energy barriers and the reaction driving forces, $\Delta E$:
\begin{equation}
    E_a = E_0 + \alpha \Delta E.
\end{equation} In our model, the \ac{BEP} coefficient $\alpha$ was $\frac{1}{2}$, and the reaction energies were derived from a bond counting model (See Supplementary Text S2). The intrinsic barriers $E_0 = E_0^{\text{hop}}$ for the terrace-diffusion are taken from the free-energy profiles (Fig.~\ref{fig:energies}C) of $0.56$~eV for Ga-polar interface. For the step descent, $E_0=E_0^{\text{down}} = E_0^{\text{hop}} + E^{ES}$, with the ES penalty obtained from the step-edge free energies (Fig.~\ref{fig:step}, value used for each edge geometry matches the corresponding metadynamics barrier). 

The rate for the third event, adsorption, is calculated directly from the net N flux ($J_{\textbf{N}}$) supplied to the interface. This flux is converted into a per-site adsorption rate, $r_\text{ad}$, by normalizing it with the surface site density, $n_s^{(f)}$, of the crystal:
\begin{equation}
    r_\text{ad} = \frac{J_\text{N}}{n_s^{(f)}}.
\end{equation}
For the $(0001)$ GaN surface, the site density is given by $n_{s}^{(0001)}=2/(\sqrt{3}a^2)$, where $a$ is the lattice constant. Using the value for GaN at 1000~K ($a=3.237$~Å), the site density is calculated to be $n_s\approx1.1\times10^{19}$~m$^{-2}$. As shown in Fig.~\ref{fig:kmc}, we conducted \ac{KMC} simulations at two different fluxes. Under the high N flux ($\sim 10^{28}$~m$^{-2}$s$^{-1}$), the rate $r_i$ is equivalent to $10^{9}$ Hz. In contrast, under the low N flux ($\sim 10^{24}$~m$^{-2}$s$^{-1}$), the rate $r_i$ is equivalent to $10^{5}$ Hz.

At each \ac{KMC} step, the algorithm first identifies the set of all possible kinetic events based on the current configuration of the surface. The rate for each potential event is calculated, and they are summed to find the total rate for the entire system $R=\sum_i r_i$. Next, a single event, $i$, is chosen from the list with a probability $p_i=r_i/R$, that is directly proportional to its rate. The simulation time then advanced by a time step, $\Delta t= -\ln \xi/R$ where $\xi$ is a random number uniformly distributed in $(0,1)$. Finally, the chosen event is executed, and the system is updated.

\paragraph*{Experimental procedures}
Ga droplets and partially nitridated Ga/GaN particles were grown in an ultra-high-vacuum \ac{MBE} with a Ga effusion cell and a radio-frequency (RF) N plasma source, as described elsewhere~\cite{liu2024influence,lu2021influence}. Substrates were amorphous SiN$_x$ TEM grids from SimPore. Prior to loading, all substrates were dipped in 5\% HF solution to remove native oxide. The substrates were held at growth temperatures for 30~min and were continuously rotating to ensure uniformity during the whole process.

Droplet epitaxy proceeded in two steps at a fixed substrate temperature of 650~$^\circ$C. First, Ga was deposited from the effusion cell to nucleate a dense array of liquid Ga droplets on the substrate surface. During which a Ga beam equivalent pressure (BEP) of $4\times10^{-7}$~Torr. Immediately afterward, nitridation was performed by igniting the RF plasma in high-purity N$_2$. The N flow rate and plasma power were held at 1~sccm and 550~W, respectively, generating a nitrogen pressure of $5\times10^{-7}$~Torr, as measured by the partial pressure of 14 amu using a residual gas analyzer. Nitridation times of 1~min and 5~min were used to obtain the “partially nitridated” Ga/GaN particles.

Additionally, sample characterizations were performed using \ac{HRTEM} and \ac{STEM}, with a Thermo Fisher Scientific Talos F200X G2 S/TEM operated at 200 kV.

\newpage



	


\clearpage 

%
\bibliography{main} 
\bibliographystyle{sciencenum}

%
%
%
%
%
%


\section*{Acknowledgments}
\textbf{Funding:}
This work was supported by the US Department of Energy, Office of Basic Energy
Sciences under Contract No. DE-SC0023222. This research used Electron Microscopy resources of the Center for Functional Nanomaterials (CFN), which is a U.S. Department of Energy Office of Science User Facility, at Brookhaven National Laboratory under Contract No. DE-SC0012704.
\textbf{Author contributions:}
Z.X. conceived and performed simulations, processed and analyzed the data, and wrote the manuscript. A.L. designed and performed experiments, processed and analyzed the data,  and wrote the manuscript. X.C. and M.L. performed the experiments and reviewed the manuscript. J.Y., R.G., and L.Q. acquired funding and resources, supervised the project, and reviewed and edited the manuscript.
\textbf{Competing interests:}
The authors declare that they have no competing interests.
\textbf{Data and materials availability:} All data needed to evaluate the conclusions in the paper are present in the paper and/or the Supplementary Materials.


\subsection*{Supplementary materials}
Supplementary Text\\
Figs. S1 to S6\\
References \textit{(66-\arabic{enumiv})} 


\newpage


\renewcommand{\thefigure}{S\arabic{figure}}
\renewcommand{\thetable}{S\arabic{table}}
\renewcommand{\theequation}{S\arabic{equation}}
\renewcommand{\thepage}{S\arabic{page}}
\setcounter{figure}{0}
\setcounter{table}{0}
\setcounter{equation}{0}
\setcounter{page}{1} 


\begin{center}
\section*{Supplementary Materials for\\ \scititle}

        Zhucong~Xi \textit{et al.}\\
\small Corresponding author: Liang Qi, qiliang@umich.edu
\end{center}

\subsection*{This PDF file includes:}
Supplementary Text S1 to S6\\
Figures S1 to S6\\


\newpage


\subsection*{Supplementary Text S1. Migration barriers on GaN clean surface}
We computed their migration barrier on the clean Ga-polar WZ$(0001)$ surface via different simulation methods. These results establish the reference for interfacial kinetics in the absence of liquid Ga and are used in the main text to highlight the catalytic effect of liquid Ga. The slab contained four Ga-N bilayers separated by $\ge 12$~Å vacuum. There is one single N adatom embedded on the surface. The bottom surface was H-passivated. In-plane supercells of $2\times4$ units (16 N and 16 Ga atoms) were used.

\begin{figure}[!ht]
    \centering
    \includegraphics[width=0.7\textwidth]{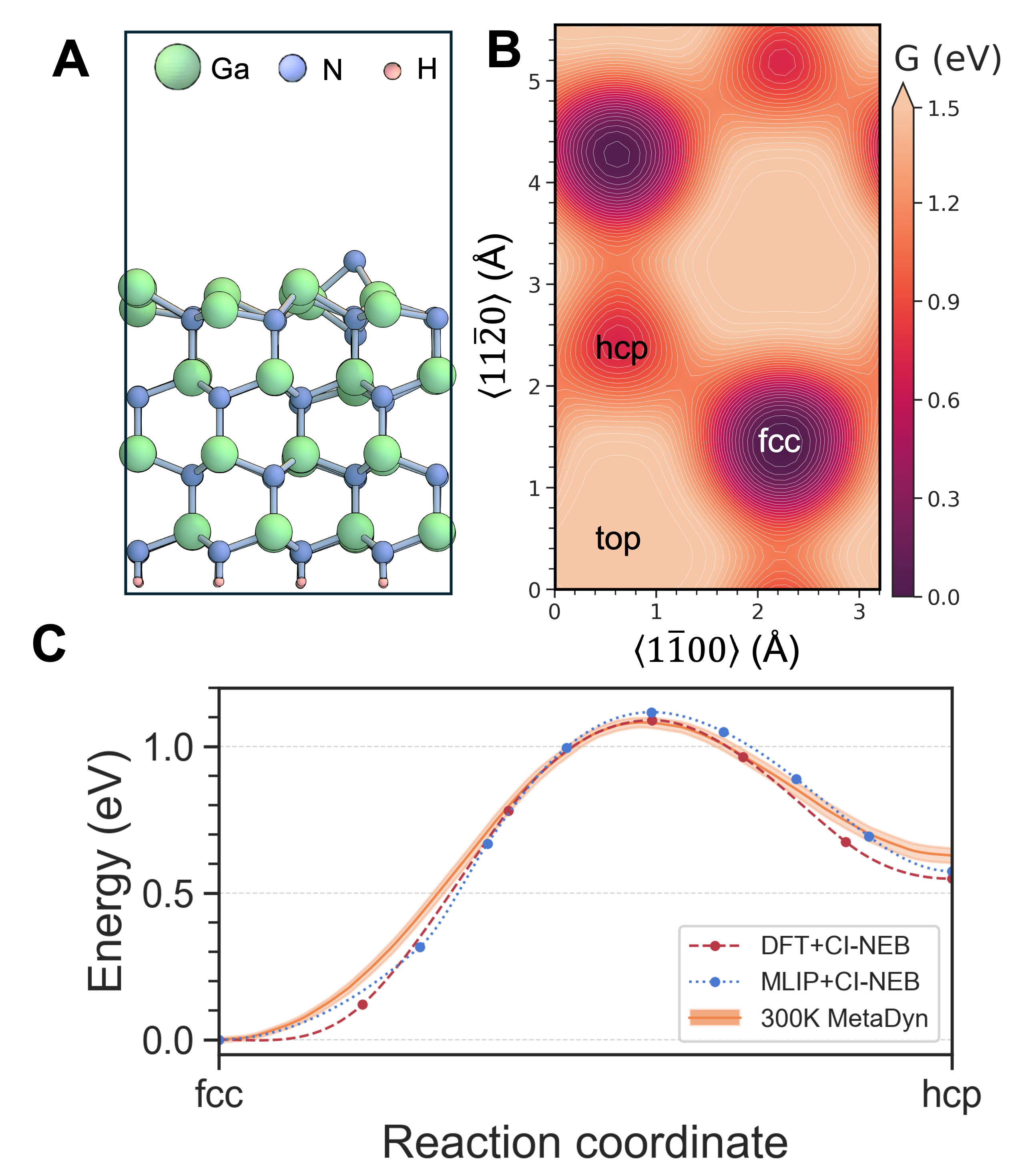}
    \caption{\textbf{Migration barriers for an N adatom at GaN clean surface.} (A): Slab model of the clean Ga-polar WZ$(0001)$ surface used for calculations; the back side is H-passivated (Ga: green, N: blue-gray, H: pink). (B) Two-dimensional free-energy surface $G(x,y)$ at 300~K from well-tempered metadynamics for an N adatom restrained to the surface. Minima occur with threefold rotational symmetry at fcc and hcp sites. Axes are along $\langle 1\bar{1}00\rangle$ and $\langle 11\bar{2}0\rangle$. (C) Minimum-energy path between hollow sites showing excellent agreement among DFT+CI-NEB (red dashed), MLIP+CI-NEB (blue dotted), and 300~K metadynamics (orange band, mean $\pm$ standard error). All methods yield a clean-surface migration barrier of $\sim$1.1~eV, much larger than the barriers at the solid-melt interfaces reported in the main text.}
    \label{fig:clean_surface} 
\end{figure}

\ac{DFT} calculations, conducted by \ac{VASP}~\cite{kresse1996efficiency,kresse1996efficient} and the \ac{VTST} package~\cite{henkelman2000climbing,henkelman2000improved}, employed the \ac{PBE} exchange-correlation functional~\cite{perdew1996generalized} with \ac{PAW} pseudopotentials~\cite{blochl1994projector}. The bottom of the slab was passivated with pseudo-H atoms, which have fractional valence electrons. The total energies for supercells of the initial and final states were converged to $10^{-6}$ eV/cell for the ionic relaxation loop and $10^{-7}$ eV for the electronic self-consistency loop, using a plane-wave cutoff energy of $550.0$~eV and Gaussian smearing of $0.05$~eV. A $3 \times 3 \times 1$ $\Gamma$-centered k-point grid was applied for the supercell with the k-spacing value of $\sim 0.20$~Å$^{-1}$. Five intermediate images were linearly interpolated between the relaxed initial and final states. The artificial spring constant was set to $5$~eV/Å$^{2}$. 

\ac{MLIP} \ac{CI-NEB} and metadynamics calculations used a larger $4\times8\times8$ slab. Initial structures from the \ac{MLIP} were converged using a Conjugate Gradient algorithm with \ac{LAMMPS}~\cite{plimpton1995fast,thompson2022lammps}. Seven middle images were linearly interpolated for \ac{CI-NEB}. Free-energy barriers at finite temperatures were obtained from well-tempered metadynamics using PLUMED~\cite{bonomi2019promoting,tribello2014plumed} plus \ac{LAMMPS}. The simulation temperature was set to 300~K with a Langevin thermostat (0.1~ps damping). We used the same two-dimensional CVs as in the interface lateral-migration calculations: the in-plane coordinates $(x,y)$ of the N adatom, defined by projections onto the two surface lattice vectors. Periodic boundary conditions were applied in CV space so that $(x,y)$ was mapped to the first unit cell during bias deposition and reweighting, enabling more efficient sampling of symmetry-equivalent minima. Well-tempered metadynamics parameters matched the main-text setup: initial hill height of 0.01~eV, Gaussian widths of 0.20~Å, bias factor of 5, and hill stride of 0.4~ps, and total biasing time of 10~ns.

All three approaches yield mutually consistent clean-surface migration kinetics. On Ga-polar WZ-GaN$(0001)$, the migration barrier is $\sim1.1$~eV along the pathway from the fcc sites to the hcp sites. This agrees with prior reports~\cite{chugh2017adsorbate}. These clean-surface barriers are $\sim$2-3 times larger than the liquid-covered interface migration free energies (0.56~eV), quantitatively supporting the Ga catalytic effects advanced in the main text.

\clearpage

\subsection*{Supplementary Text S2. N pair‐interaction at the interface}
To quantify the effective interaction between two nitrogen adatoms at the GaN(\textit{s})-Ga(\textit{l}) interface, we computed one-dimensional free energy using well-tempered metadynamics with the same \ac{PFP}, $T=1000$~K, 1~fs timestep, and a Langevin thermostat (0.1~ps damping) as in the main-text simulations. Ga-polar WZ$(0001)$ and N-polar WZ$(000\bar{1})$ slabs (768-atom GaN plus a 500-atom Ga melt) were employed. Two N adatoms were then introduced at the solid-melt interface. The systems were pre-equilibrated for 5~ps. The \ac{CV} was the in-plane distance separation distance of two N adatoms, biased with Gaussian hills (initial height 0.01~eV, width 0.10~Å, stride 0.4~ps, bias factor 5) for 10~ns. To isolate lateral interactions, each N was softly restrained along $z$ to the first adsorption well of its respective polarity (identified from the single-adatom vertical free energy as shown in Fig.~\ref{fig:order}E and F), preventing drift into the bulk liquid or desorption. 

\begin{figure}[!htbp]
    \centering
    \includegraphics[width=1.0\linewidth]{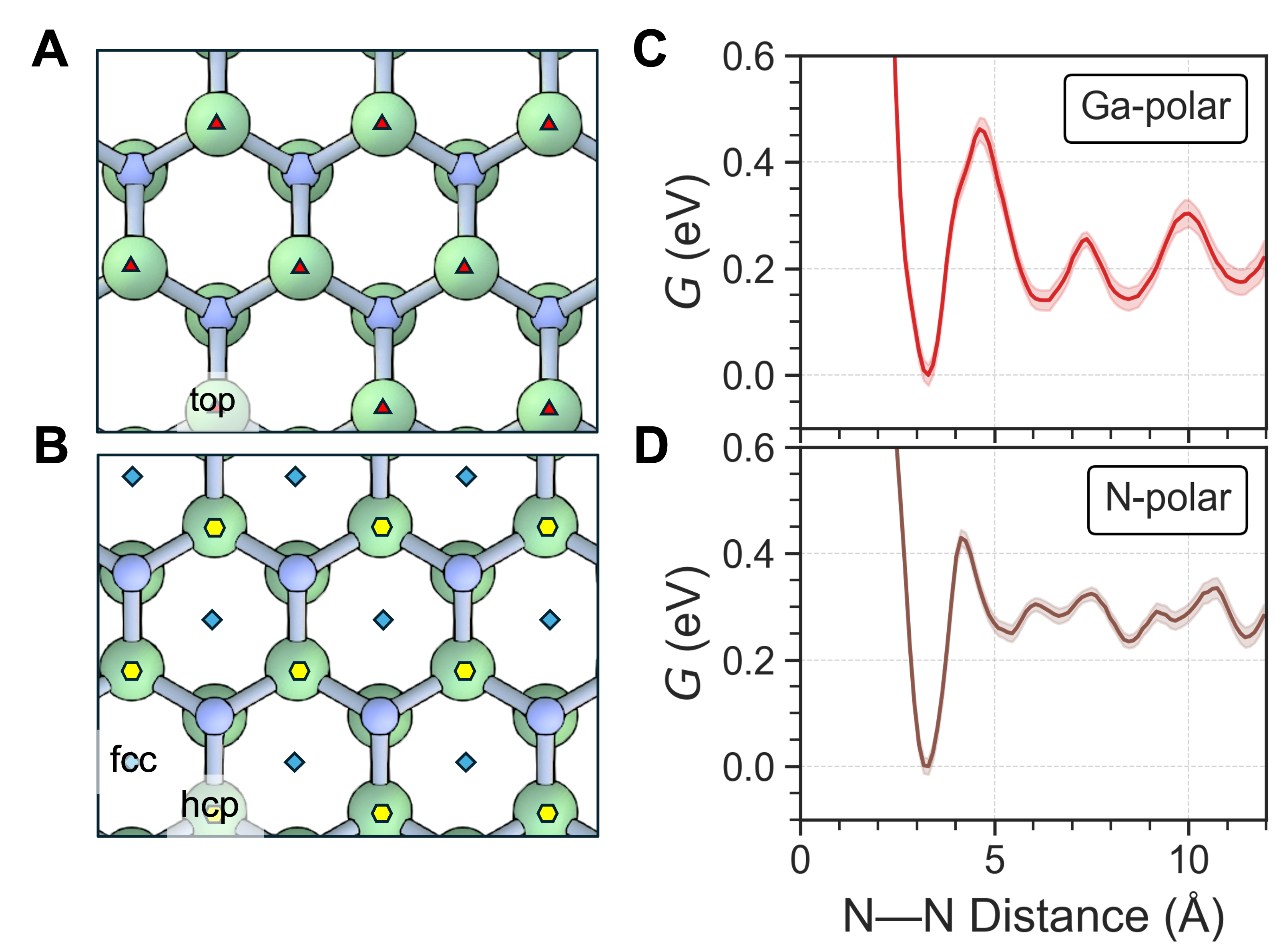}
    \caption{\textbf{Pair‐interaction free energies of N adatoms at Ga(\textit{l})-GaN(\textit{s}) interfaces.} (A)-(B): Top views of the Ga-polar (A) and N-polar (B) terraces showing the preferred adsorption sites for reference N adatoms (Ga-polar: top; N-polar: fcc/hcp). (C)-(D): Interaction free energy $G$ between two N adatoms as a function of their in-plane separation distance, obtained by well-tempered metadynamics at $T=1000$~K. Both terminations exhibit strong short-range repulsion and weak oscillatory coupling at larger separations. Shaded bands indicate mean $\pm$ standard error from block-averaged reweighting.}
    \label{fig:pair}
\end{figure}

Fig.~\ref{fig:pair} summarizes the two-adatom metadynamics used to describe the interactions between N adatoms at the interface. Figs.~\ref{fig:pair}A and B show atomistic snapshots of preferred adsorption sites for the Ga-polar WZ$(0001)$ and N-polar WZ$(000\bar{1})$ interfaces, respectively (identified from the single-adatom lateral free energy as shown in Fig.~\ref{fig:energies}). As presented in Figs.~\ref{fig:pair}C and D. In both cases, the free energy profiles $G$ exhibit a steep short-range repulsion that forbids co-occupation of the same site, a pronounced minimum at
$\sim$3.2~Å, corresponding to the first nearest neighbor distance, and damped oscillations at larger distances arising from the periodic spacing of adsorption sites. In both cases, the free-energy profile shows a difference of about $\sim$0.2 eV, corresponding to the bonding energy between two N adatoms at the interface, denoted as $E_b^{\text{N-N}}$. This value was then used in the bond-counting model to estimate the driving force for N migration, expressed as
\begin{equation}
\Delta E = \Delta n \cdot E_b^{\text{N-N}}
\end{equation}
Here, $\Delta n$ is the number of first-nearest N-N bond changes before and after the N adatom hop at the interface. This formulation directly links the change in bond count to the corresponding migration driving force. Thus, the bond-counting model provides a simple yet quantitative way to capture how N-N bonding influences migration.

\clearpage

\subsection*{Supplementary Text S3. Diffusivity of N in liquid Ga}
We determined the tracer diffusivity of N in liquid Ga, from equilibrium \ac{MD} using the same \ac{MLIP}. To do so, we simulated 800 Ga atoms in liquid plus a single N probe atom in a cubic periodic cell at sampled temperatures of 900, 1000, 1100, and 1200~K. The system was equilibrated for 20~ps in the isothermal-isobaric (NPT) ensemble using the Nosé-Hoover barostat with damping of 1~ps and the Langevin thermostat with damping of 0.1~ps. After equilibration, both the thermostat and the barostat were removed, and the simulation was run in the microcanonical (NVE) ensemble for 200~ps.

\begin{figure}[!ht]
    \centering
    \includegraphics[width=1.0\linewidth]{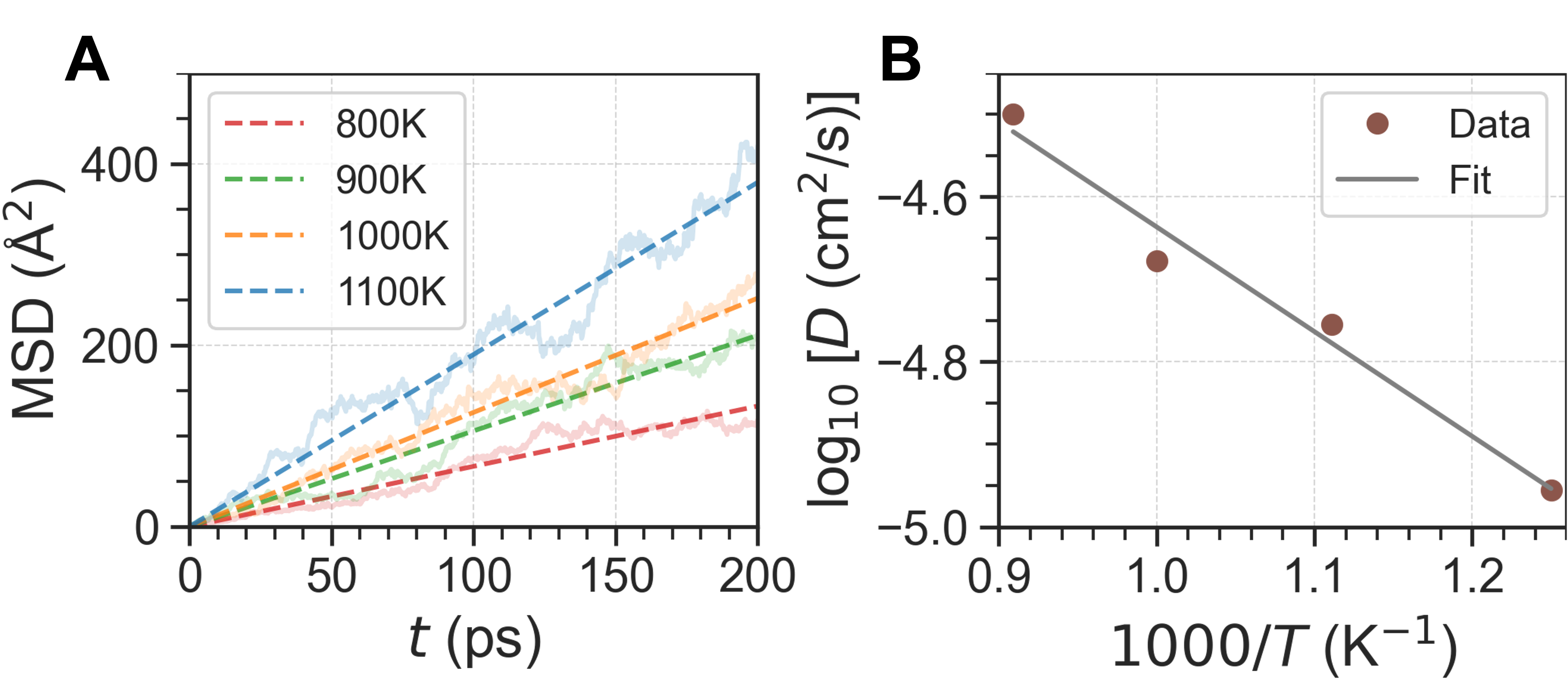}
    \caption{\textbf{Diffusivity of N in liquid Ga.} (A) Mean-squared displacement (MSD) of a single N probe in molten Ga at 800-1100K (solid curves). Dashed lines are linear fits in the diffusive regime used with the Einstein relation. (B) Arrhenius representation of the diffusivity, $\log_{10} D$ (cm$^2$s$^{-1}$) versus $1000/T$; points are MD values and the solid line is the best linear fit, whose slope yields the activation energy and intercept the prefactor. The resulting $D(T)$ is used in the diffusion-limited growth model.}
    \label{fig:diffusivity}
\end{figure}

During the simulation, we measured the \ac{MSD} of the N probe:
\begin{equation}
\langle [\Delta R(t)]^2 \rangle
=\frac{1}{n_{\text{probe}}} \sum_{i=1}^{n_{\text{probe}}} \left[ \pmb{R}_i(t) - \pmb{R}_i(0) \right]^{2}.
\end{equation}
Here, $n_{\text{probe}}$ is the number of probe atoms, which equals 1 in our setup. $\pmb{R}_i(i)$ is the position of atom $i$ at $t$ time. The MSD was accumulated along each temperature as shown in Fig.~\ref{fig:diffusivity}A. The diffusivity of N was obtained from the Einstein relation for three-dimensional diffusion,
\begin{equation}
D=\frac{1}{6}\frac{d}{dt}\langle [\Delta R(t)]^2 \rangle,
\end{equation}
where $D$ was computed by first making a linear approximation to the \ac{MSD}, as shown in the dashed line Fig.~\ref{fig:diffusivity}A.

An Arrhenius analysis was performed by fitting the diffusivity as a function of temperature, $D(T)$:
\begin{equation}
D(T)=D_0\exp\left(-\frac{E_D}{k_{\rm B}T}\right).
\end{equation}
This is equivalently a linear regression of $\log D$ versus $1/T$, as shown in Fig.~\ref{fig:diffusivity}B. Points in Fig.~\ref{fig:diffusivity}B are the values obtained from the \ac{MD} simulations, and the solid line is the least squares linear fit. The slope yields the activation energy $E_D$ and the intercept gives the prefactor $D_0$. The resulting temperature-dependent diffusivity $D(T)$ is used directly in the diffusion-limited growth model in the main text.

\clearpage

\subsection*{Supplementary Text S4. Solubility of N in liquid Ga}
\begin{figure}[!ht]
    \centering
    \includegraphics[width=1.0\linewidth]{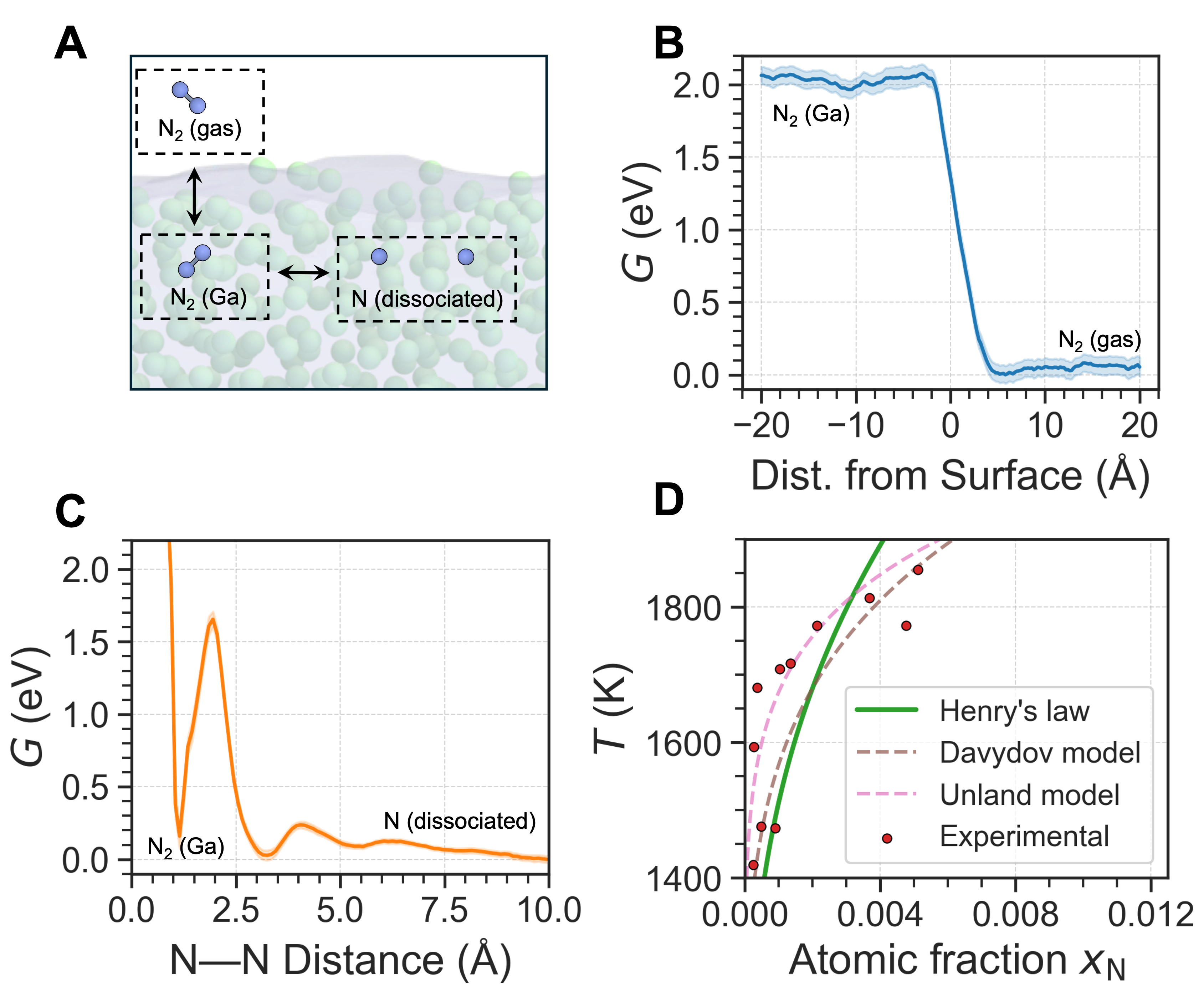}
    \caption{\textbf{Solubility of N in Ga liquid from metadynamics simulations.} (A): Conceptual illustration of N$_2$ solvation with an N$_2$ molecule in the gas phase, an N$_2$ molecule solvated in Ga, and two dissociated N atoms in the Ga liquid. (B): Solvation free-energy profile $G(z)$ at 1000~K as a function of the distance of N$_2$ from the liquid surface, showing a $\sim$2~eV to transfer N$_2$ into Ga. (C): Free-energy profile for N$_2$ dissociation in Ga versus N-N separation distance, with a $\sim$1.7~eV barrier and an dissociation energy of $\sim$0.15~eV per N pair. (D): Predicted nitrogen atomic fraction $x_{\text{N}}$ as a function of temperature using the Henry's law expression derived from our  free energy (green), compared with two empirical CALPHAD models (Davydov~\cite{davydov1999thermodynamic,davydov2001thermodynamic} and Unland~\cite{unland2003thermodynamics}; dashed) and representative experimental points (red circles)~\cite{grzegory1995iii,thurmond1972equilibrium,logan1972heteroepitaxial,madar1975high,karpinski1984equilibrium}. All approaches indicate extremely low N solubility in Ga.}
    \label{fig:solubility}
\end{figure}

We quantified the equilibrium solubility of nitrogen in liquid Ga through a two-step free energy calculation evaluated by well-tempered metadynamics at $1000$~K using the \ac{MLIP} described in the main text. The conceptual illustration of this two-step method is shown in Fig.~\ref{fig:solubility}A. The first step is the transfer of an N$_2$ molecule from the gas phase into the Ga melt, and the second step is dissociation of the solvated N$_2$ molecule. 

For the first step:
\begin{equation}
\text{N}_2(\text{gas}) \leftrightarrow \text{N}_2(\text{Ga~liquid}),
\end{equation}
the solvation free energy $\Delta G_{\text{solv}}^{\text{N}_2}$ was computed with a \ac{CV} defined as the distance between the center of mass of the Ga liquid slab and that of N$_2$. The simulation cell contained 800 Ga atoms and one N$_2$ molecule. A Lennard-Jones wall at the bottom of the slab stabilized the liquid and prevented drift. The system was pre-equilibrated for 10~ps with a Langevin thermostat (0.1~ps damping). Gaussian hills of width 0.30~Å and initial height 0.3~eV were deposited every 0.4~ps with a bias factor of 20 for a total of 10~ns. The resulting profile, as shown in Fig.~\ref{fig:solubility}B, shows a $\sim2.0$~eV penalty per molecule to transfer N$_2$ into the liquid, indicating very unfavorable solvation at these conditions.

For the second step:
\begin{equation}
\text{N}_2(\text{Ga~liquid}) \leftrightarrow 2\text{N}(\text{Ga~liquid}),
\end{equation}
the dissociation free energy, $\Delta G_{\text{diss}}^{\text{N}_2}$, was evaluated in a fully periodic box containing 800 Ga atoms and one N$_2$. After 10~ps equilibration at 1000~K using a Langevin thermostat and Nosé-Hoover barostat, we biased the N-N bond length as the CV. Gaussian width was 0.10~Å, initial height 0.04~eV, bias factor 10, and hills were deposited every 0.4~ps for 10~ns. The free-energy profile, as shown in Fig.~\ref{fig:solubility}C, exhibits a dissociation barrier of $\sim$1.6-1.7~eV and a dissociation free energy of $\sim$0.15~eV per molecule at 1000~K.

Combining the two steps yields the standard free energy for producing dissolved atomic N from gas-phase N$_2$:
\begin{equation}
\frac{1}{2}\text{N}_2(\text{gas}) \leftrightarrow \text{N}(\text{Ga~liquid}),
\end{equation}
with the reaction energy $\Delta G_{\text{solv}}^{\text{N}}(T)=\frac{1}{2}\left(\Delta G_{\text{solv}}^{\text{N}_2}(T)+\Delta G_{\text{diss}}^{\text{N}_2}(T)\right)\approx$ 0.85~eV/atom at 1000~K. At equilibrium, it requires
\begin{equation}
    \frac{1}{2}\mu^{\text{gas}}_{\text{N}_2}(T,p_{\text{N}_2})=\mu^{\text{liquid}}_{\text{N}}(T,x_{\text{N}})\,
\end{equation}
with $\mu^{\text{gas}}_{\text{N}_2}=\mu^{\circ,\text{gas}}_{\text{N}_2}(T)+k_BT\ln(p_{\text{N}_2}/p^\circ)$ and $\mu^{\text{liquid}}_{\text{N}}=\mu^{\circ,\text{liquid}}_{\text{N}}(T)+k_BT\ln x_{\text{N}}$. Here, $\mu^{\circ,\text{gas}}_{\text{N}_2}(T)$ is the standard chemical potential of ideal N$_2$ at the standard pressure of 1~bar, $\mu^{\circ,\text{liquid}}_{\text{N}}(T)$ is the standard chemical potential of dissolved atomic N in the dilute Ga. Solving for the atomic-N molar concentration yields Henry's law:
\begin{equation}
x_{\text{N}_2}(T,p_{\text{N}_2}) = \exp\left(\frac{\frac{1}{2}\mu^{\circ,\text{gas}}_{\text{N}_2}- \mu^{\circ,\text{liquid}}_{\text{N}}}{k_B T}\right) \left(\frac{p_{\text{N}_2}}{p^\circ}\right)^\frac{1}{2}=\exp\left(-\frac{\Delta G_{\text{solv}}^{\text{N}}}{k_B T}\right) \left(\frac{p_{\text{N}_2}}{p^\circ}\right)^\frac{1}{2}
\label{eq:henry}
\end{equation}

Using $\Delta G_{\text{solv}}^{\text{N}}\approx0.85$~eV at 1000~K, at which $k_BT\!\approx\!0.086$~eV, gives the prefactor
$\exp[-\Delta G_{\text{solv}}^{\text{N}}/k_BT]\!\approx\!5\times10^{-5}$; thus
$x_{\mathrm N}\propto \sqrt{p_{\mathrm{N_2}}}$ is extremely small even near 1~bar. Davydov \textit{et al.}~\cite{davydov1999thermodynamic,davydov2001thermodynamic} and Unland \textit{et al.}~\cite{unland2003thermodynamics} evaluated the thermodynamic assessment of the Ga-N system~\cite{grzegory1995iii,thurmond1972equilibrium,logan1972heteroepitaxial,madar1975high,karpinski1984equilibrium} and gave empirical CALPHAD models. Fig.~\ref{fig:solubility}D compares the $x_{\text{N}}(T)$ curve from Eq.~\ref{eq:henry} with Davydov and Unland fit, as well as experimental data. The comparison is limited to 1400-1900~K, which is the temperature window analyzed in those models and covered by available measurements.


\clearpage

\subsection*{Supplementary Text S5. Derivation for the steady-state equation}
We treat the interfacial incorporation of N as diffusion driven by the concentration gradient and the free energy gradient. The total N flux in the direction perpendicular to the growth plane is given:
To derive the steady-state solution for the N adatom flux, $J$, we begin with the master equation, which is a modified form of Fick's first law incorporating a chemical potential gradient:
\begin{equation}
    J_{\text{N}} = -D \frac{dc(z)}{dz} - \frac{Dc(z)}{k_B T} \frac{dG(z)}{dz}
\end{equation}
where $D$ is the diffusivity, $c(z)$ is the concentration, $G(z)$ is the chemical potential (free energy), $k_B$ is the Boltzmann constant, and $T$ is the temperature. Under steady-state conditions, the flux $J_{\text{N}}$ is constant. We can rearrange this first-order linear ordinary differential equation into the standard form:
\begin{equation}
\frac{dc(z)}{dz} +\frac{c(z)}{k_B T} \frac{dG(z)}{dz}  = -\frac{J_{\text{N}}}{D}    
\end{equation} 
This equation can be solved using an integrating factor $\exp\left(\frac{G(z)}{k_B T}\right)$:
\begin{equation}
     \exp\left(\frac{G(z)}{k_B T}\right) \frac{dc}{dz} + \exp\left(\frac{G(z)}{k_B T}\right) \frac{c(z)}{k_B T} \frac{dG}{dz} = - \frac{J_{\text{N}}}{D} \exp\left(\frac{G(z)}{k_B T}\right)
\end{equation}
Thus, our equation becomes:
\begin{equation}
    \frac{d}{dz}\left(c(z) \exp\left(\frac{G(z)}{k_B T}\right)\right) = -\frac{J_{\text{N}}}{D} \exp\left(\frac{G(z)}{k_B T}\right)
\end{equation}
We now integrate both sides with respect to $z$ from the interface at $z=0$ to a position $z=L$ far into the liquid:
\begin{equation}
 \int_0^L \frac{d}{dz}\left(c(z) \exp\left(\frac{G(z)}{k_B T}\right)\right) dz = -\frac{J_{\text{N}}}{D} \int_0^L \exp\left(\frac{G(z)}{k_B T}\right) dz
\end{equation}
Finally, applying the fundamental theorem of calculus and rearranging the terms in the numerator gives the expression for the steady-state flux as:
\begin{equation}
    J_{\text{N}} = D \left( \frac{ c(0) \exp \left( \frac{G(0)}{k_B T} \right) - c(L) \exp \left( \frac{G(L)}{k_B T} \right)}{\int_0^{L} \exp \left( \frac{G(z)}{k_B T} \right) dz} \right)
\end{equation}
The sign convention takes $+z$ from solid to liquid; thus, a negative $J_{\text{N}}$ corresponds to net flux from liquid toward the solid (growth).

\clearpage

\subsection*{Supplementary Text S6. Crystallographic statistics of partially nitridated GaN/Ga particles}
\begin{figure}[!ht]
    \centering
    \includegraphics[width=1.0\linewidth]{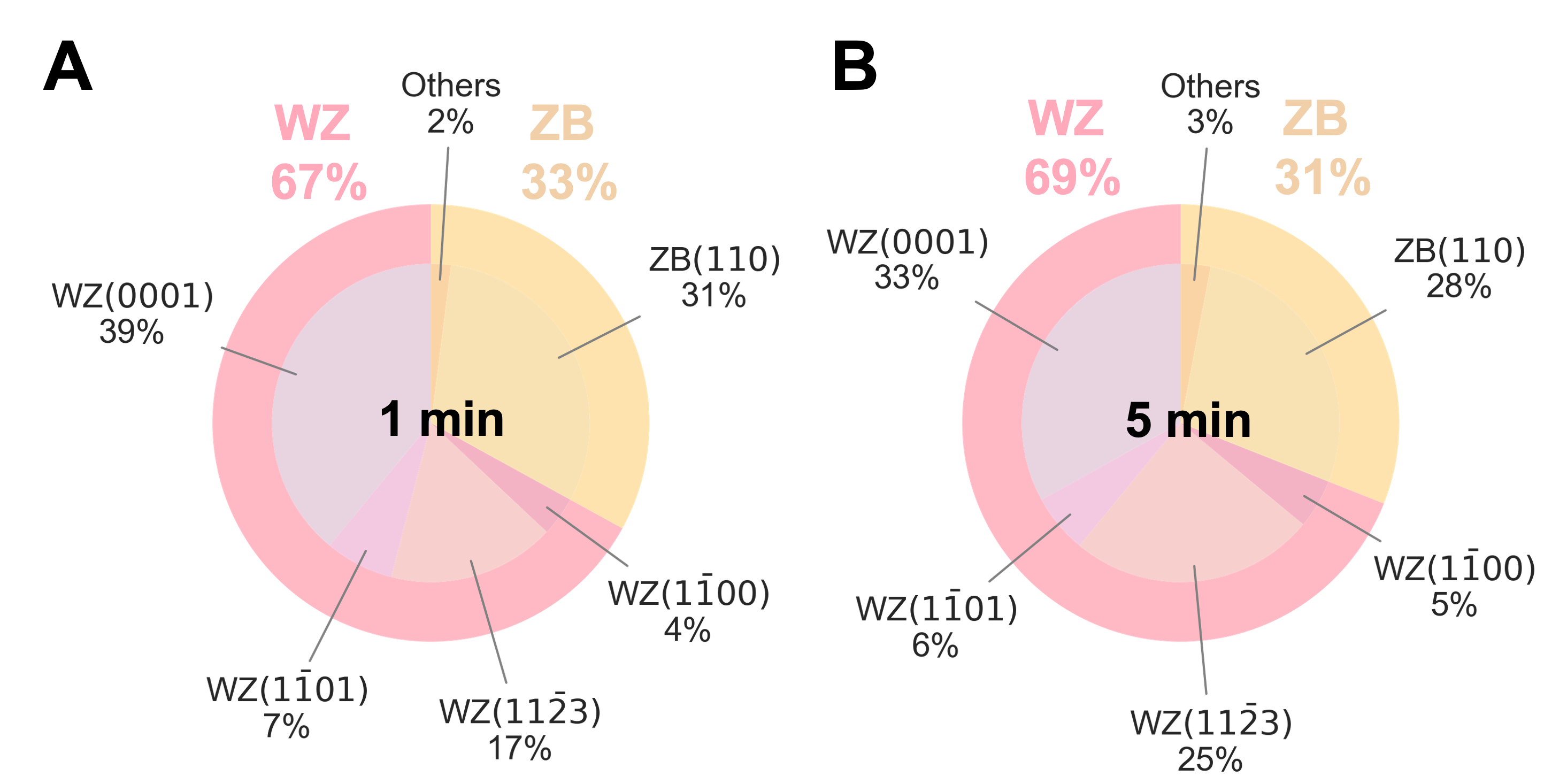}
    \caption{\textbf{The crystallographic information and GaN radius of the 1 min and 5 min partially nitridated GaN/Ga particles.} Distribution of crystal plane appearances in 1 min (A) and 5 min (B) partially nitridated GaN/Ga particles.}
    \label{fig:crystallographic}
\end{figure}

\begin{figure}[!ht]
    \centering
    \includegraphics[width=0.8\linewidth]{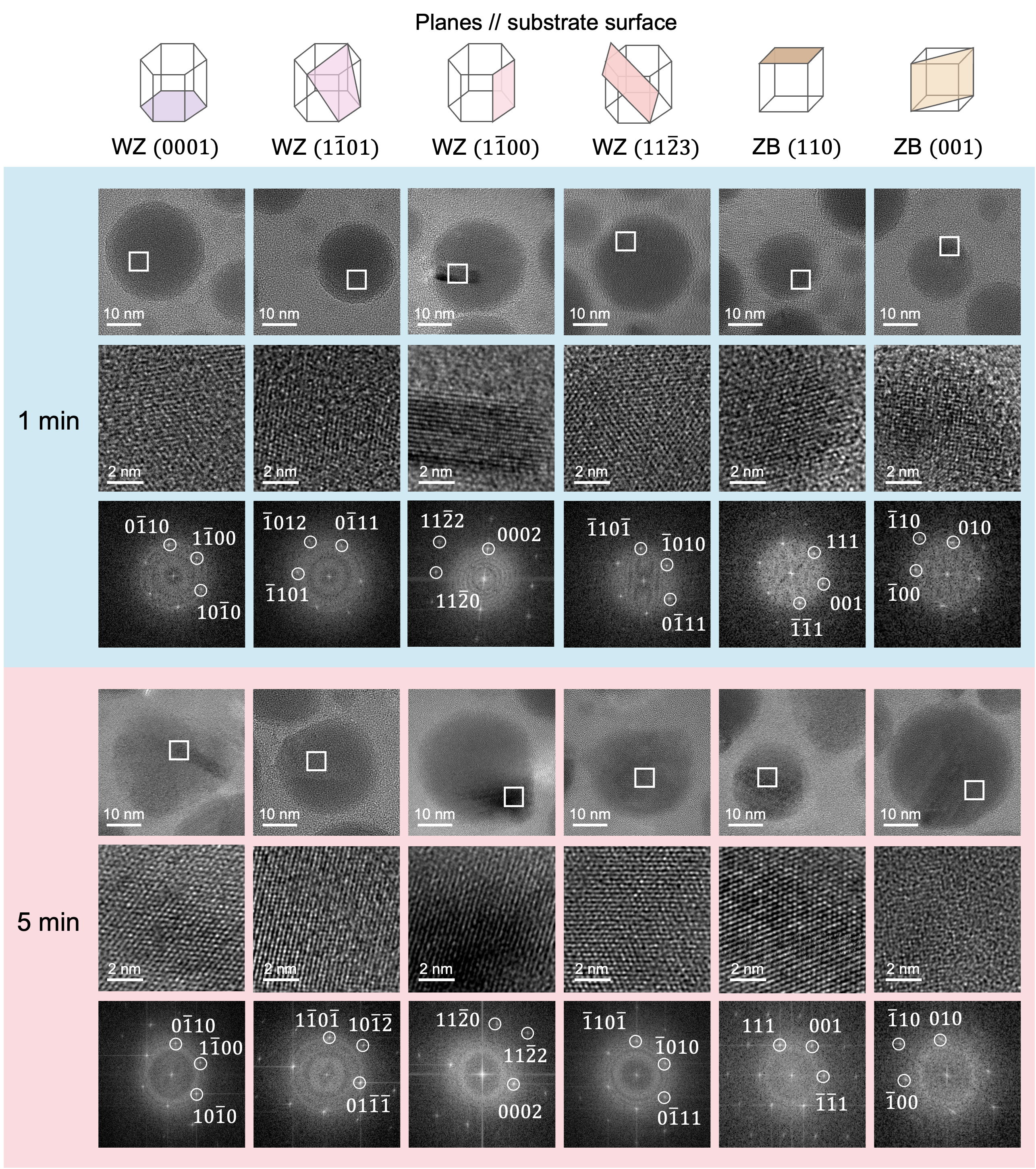}
    \caption{\textbf{Identification of the crystal planes of individual GaN/Ga particles.} Panels are grouped by nitridation time (top, blue: 1~min; bottom, pink: 5~min). For each crystallographic assignment, the top row shows a HAADF-STEM overview of a representative particle with a white square marking the region of interest; the middle row shows a HRTEM zoom of that boxed area; and the bottom row shows the 2D FFT of the HRTEM image with indexed reflections.}
    \label{fig:more_tem}
\end{figure}

We indexed more than 200 partially nitridated GaN/Ga particles after 1 and 5~min N$_2$ plasma exposure. The crystallographic statistics are summarized in Fig.~\ref{fig:crystallographic}, and the identification of the crystal planes of individual GaN/Ga particles is presented in Fig.~\ref{fig:more_tem}. Regardless of the nitridation time, the \ac{WZ} polytype has roughly one third (67\% at 1~min and 69\% at 5~min), with the remainder of \ac{ZB}.Within WZ, the most frequent orientations are the $(0001))$ and $(11\bar{2}3)$, followed by prismatic/semipolar facets $(1\bar100)$ and $(1\bar{1}01)$. Within ZB, $(110)$ is predominant, and others facets, like $(001),$ only have a small amount. The close agreement between the 1 and 5~min datasets indicates that polytype and facet selection are largely determined at nucleation, with subsequent nitridation primarily increasing particle size rather than altering crystallography.

\end{document}